\documentclass{aa}  

\usepackage{graphicx}
\usepackage{xcolor}
\usepackage{txfonts}
\usepackage[colorlinks=true, citecolor=blue]{hyperref}
\usepackage{array}
\usepackage{tabularx}
\usepackage{amsmath}
\usepackage{multirow}
\usepackage{multicol}
\usepackage{booktabs} 
\usepackage{natbib}
\usepackage{mathtools}
\usepackage{listings}
\usepackage{verbatim}
\usepackage[italicdiff]{physics}

\bibpunct{(}{)}{;}{a}{}{,} 

\usepackage{comment}
\usepackage{color}
\usepackage[normalem]{ulem}

\newcommand{\vx}{\mathbf{x}}
\newcommand{\vv}{\mathbf{v}}

\begin{document} 

   \title{ Are heuristic switches necessary to control dissipation in modern smoothed particle hydrodynamics?}
   \titlerunning{Are heuristic switches necessary in modern SPH?}

   \author{Domingo García-Senz\inst{1,2}
          \and
          Rubén M. Cabezón\inst{3,4}}

   \institute{Departament de Física. Universitat Politècnica de Catalunya (UPC). Av. Eduard Maristany 16, 08019 Barcelona, Spain \email{domingo.garcia@senior.upc.edu} \and Institut d'Estudis Espacials de Catalunya (IEEC), 08860 Castelldefels (Barcelona), Spain
    \and
    Center for Scientific Computing - sciCORE, University of Basel, Klingelberstrasse 61, 4056 Basel, Switzerland \email{ruben.cabezon@unibas.ch}
    \and
    Center for Data Analytics - CEDA, University of Basel, 4056 Basel, Switzerland}

   \date{XXX}

 
  \abstract
  {Artificial viscosity is commonly employed in smoothed particle hydrodynamics (SPH) to model dissipation in hydrodynamic simulations. However, its practical implementation today relies, in many cases, on complex numerical switches to restrict its application to regions where dissipation is physically warranted, such as shocks. These switches, while essential, are imperfect and can introduce additional numerical noise.} 
   {We investigated an efficient shock capture scheme for SPH that does not rely on artificial viscosity switches. The advantages of the proposed scheme have been validated  through a  representative number of  test cases.}
   {Recent studies have proposed that subtracting the linear component of the velocity field can suppress spurious dissipation in shear-dominated regions. Building on this idea, we implemented a velocity-reconstruction technique that removes the bulk linear motion from the local velocity field and uses the Balsara correction to modulate the dissipation.}
   {The  methodology presented here yields a balanced dissipation scheme that performs well across a range of regimes, including subsonic instabilities, shear flows, and strong shocks. We demonstrate that this approach yields improved accuracy and lower spurious dissipation, compared to the reference viscosity switch used in this work.}
   {}


   \maketitle
%

\section{Introduction}

Smoothed particle hydrodynamics (SPH)  is a Lagrangian method widely used to simulate fluid flows,  originally introduced by \cite{gingold77}
and \cite{lucy77}. It is particularly useful in astrophysical contexts where complex geometries and free surfaces are common. Despite its versatility and continued development, SPH still faces challenges in certain areas; notably, in the accurate treatment of dissipation.

Unlike grid-based methods, which typically employ Riemann solvers \citep{godunov59} to capture fluid discontinuities, SPH relies primarly on artificial viscosity (AV) formulations\footnote{There have also been proposals to adapt Riemann solvers for use in SPH \citep{inu02,mon1997,cha03, cha2010}, representing a promising parallel line of development \citep{rosswog2024}.}. The theoretical foundations of the standard AV scheme have long been established \citep{mon1983,mon1992}. However, practical applications of the method still require additional refinements in cases where the required algorithm needs to be sufficiently versatile to handle different fluid regimes.

One of these refinements has been the introduction of corrective factors (AV limiters) that suppress viscosity in strongly shearing regions, thereby reducing dissipation in vorticity-dominated flows and facilitating the growth of hydrodynamic instabilities \citep{bal95}. A widely adopted and often more effective strategy employs time-dependent dissipation coefficients (i.e., switches), which remain low by default, rising rapidly when a shock is detected and relaxing back to their nominal values once the shock has passed \citep{morris1997,mon2005}. The different implementations of those time-dependent AV parameters primarily differ in the design and responsiveness of the shock detector and the relaxation law \citep{cul10, read2012, ros15, rosswog2020}. 

Although artificial-viscosity switches (AVSWs) represent a major advance and do indeed reduce unwanted dissipation, they are not without drawbacks \citep[however, see][for an entropy-steered variant]{rosswog2020}. In post-shock regions, oscillations in the time-dependent dissipation coefficients can induce excessive numerical noise and spuriously increase entropy. Moreover, for certain problems, such as instability growth and subsonic turbulence, the reported convergence rates are often lower than those achieved with mesh-based methods \citep{bauer2012}. 

Recently, \cite{fro17} proposed the use of slope-limited reconstruction (SLR) within the artificial dissipation terms to suppress spurious viscosity. Conceptually, SLR acts as a local filter that subtracts the linear component of the velocity field, preventing bulk shear from feeding the dissipative operator. The resulting algorithm is not derived from first principles and is therefore heuristic in nature as well. However, unlike AVSW, it is purely local and instantaneous: it involves no time derivatives and does not require prescribing a working range for the viscosity coefficient $\alpha$. In previous studies, SLR has been combined with AVSW \citep{fro17,rosswog2020,rosswog_diener2025,cabezon2025} with good effect. Nevertheless, several tests indicate that further gains are achieved when AVSWs are omitted altogether \citep{sandnes2025}. Similarly, \cite{price2019} explored switch-free dissipation and reported that SLR compares favorably to switch-based schemes, albeit only for 1D test problems. The prospect of attaining high accuracy without AV switches is noteworthy and given the ubiquity of these terms in SPH, it warrants closer examination. 

 On another note, SLR schemes and, in general, algorithms based on total variation diminishing (TVD) approaches have been criticized for not being automatically consistent with the entropy inequality \citep{yee1983}. In SLR schemes, dissipation is not guaranteed to be positive-definite by construction \citep{price2020}, which may cause local violations of the second principle of thermodynamics. There are ways to modify TVD to be consistent with the entropy condition in 1D \citep{harten1976, osher1984}, but this remains an open question in 3D. However, as discussed in Sect.~\ref{sec:optimalAV}, we have not seen anomalous entropy behavior in any of the tests performed when SLR was used\footnote{To be precise, both the SLR and the AV-switches showed slight entropy inversions in noise-dominated regions and around contact discontinuities when these were not conveniently smoothed (e.g., in the shock-tube test).}. Other situations where switches can potentially outperform Godunov-SPH and SLR methods (specifically, 1D linear-wave evolution) have been discussed in \cite{puri2014a}.

In this work, we carried out a suite of SPH simulations using a SPH formulation distinct from that of \cite{sandnes2025}, with two goals: (i) to independently validate the performance of the slope-limited reconstruction (SLR) technique; and (ii) to refine it where it can prove beneficial. Our test suite included several benchmarks considered by \cite{sandnes2025} as well as additional problems that probe shear- and turbulence-dominated regimes. We have paid particular attention to turbulent flows, where excessive artificial dissipation is especially detrimental.

This article is organized as follows. Section~\ref{sec:sphflavor} summarizes the different dissipation schemes currently used to handle dissipation in, as well as our proposed methods. Section~\ref{sec:tests} presents various tests comparing the performance of these schemes. Section~\ref{sec:turbulence} is entirely dedicated to the study of subsonic turbulence and how the dissipation schemes influence SPH's ability to resolve it. The methods that offer the best balanced results across all tests are ranked in Sect.~\ref{sec:optimalAV}, where the potential disadvantages of SLR schemes are also outlined and discussed. 
Finally, in Sect.~\ref{sec:conclusions}, we discuss our main findings, along with the recommended shock-capturing setting and future prospects. 
 
\section{SPH flavor}
\label{sec:sphflavor}
Most of the test cases described in Sect.~\ref{sec:tests} were carried out with the SPH code SPHYNX \citep{cabezon2017}. Only the turbulence simulations were performed with SPH-EXA \citep{cavelan2020, keller2023}, due to its scalability and native GPU acceleration. This enabled higher-resolution, time- and resource-intensive runs to be executed substantially faster. Both codes incorporate the same improved algorithms for gradient estimation \citep{garciasenz2012,cabezon2012,ros15}, kernel interpolation \citep{cabezon2008,cabezon_mixedsincs_2024}, and an accurate partition of unity \citep{garciasenz2022}, while the equation of energy is integrated consistently to describe thermal evolution (i.e., grad-h terms are included). These ingredients are fully consistent with the Lagrangian nature of SPH, yielding a conservative scheme. The remainder of this section focuses on the construction and implementation of dissipation in artificial viscosity-based SPH schemes.

\subsection{Artificial viscosity schemes}

As in other modern SPH formulations \citep{price18,rosswog2020magma, Wadsley17}, SPHYNX and SPH-EXA handle dissipation with the classical artificial viscosity (AV) scheme \citep{mon1992,mon1997}. In particular, \cite{mon1997} defines the pairwise AV term  $\Pi_{ab}$ in analogy to a Riemann problem,

\begin{equation}
        \Pi_{ab}=-\frac{\alpha}{2}\frac{v_{ab}^{sig}\omega_{ab}}{\bar\rho_{ab}}\,,
        \label{eq:AV_mon97}
    \end{equation}

\noindent with

\begin{equation}
\omega_{ab}={\bf v}_{ab}\cdot \hat {\bf r}_{ab}
\label{eq:omega}
,\end{equation}

\noindent where ${\bf v}_{ab} = {\bf v}_a-{\bf v}_b$ and $\hat {\bf r}_{ab}$ is the unit vector joining particles $a$ and $b$. Here, 

\begin{equation}
v_{ab}^{sig}= c_a+c_b-\gamma \omega_{ab}\,
\label{eq:signalvel}
\end{equation}

\noindent is a characteristic signal velocity at which the information propagates between particles $a$ and  $b$, and $\gamma= 3$. Equations \ref{eq:AV_mon97} and \ref{eq:signalvel}  can be loosely related to the widely used expression of \cite{mon1992}, namely, 

\begin{equation}
    \Pi_{ab}=
\begin{cases}
\frac{-\alpha \bar c_{ab}\mu_{ab} + \beta \mu_{ab}^2}{\bar \rho_{ab}}& \text{for~~}{\bf v}_{ab}\cdot {\bf r}_{ab} < 0,\\
0 &  \text{for~~}{\bf v}_{ab}\cdot {\bf r}_{ab}\ge 0,
\end{cases}
\label{eq:AV_classic}
\end{equation}

\noindent
where 
\begin{equation}
\mu_{ab}= \frac{\bar h_{ab} {\bf v}_{ab}\cdot {\bf {r}_{ab}}}{{\bf r}_{ab}^2+\eta^2 }\,.
\end{equation}

Neglecting the regularization, $\eta$, and taking $\frac{\bar h_{ab}}{\vert {\bf r}_{ab}\vert} =1$ yields $\mu_{ab}\approx\omega_{ab}$; thus, comparing Eqs.\ref{eq:AV_mon97} and \ref{eq:AV_classic} gives $\beta=\gamma\alpha /2$. For $\gamma=3$, we have $\beta=\frac{3}{2} \alpha$. However, a more frequent choice is $\beta=2\alpha$, which better reduces post-shock oscillations in strong shocks. In the continuum limit, it has been shown that the linear part of AV in Eq.~\ref{eq:AV_classic} is equivalent to the shear and bulk viscosities in the Navier-Stokes equations \citep{mon2005}. Thus, the AV is not as "artificial" as it might seem after all. 
 
Modern schemes handle $\alpha$ and $\beta$ independently, since $\alpha$ is often allowed to vary per particle and in time during calculations, while $\beta$ is kept constant. For example, \cite{Wadsley17} considered

\begin{equation}
    \Pi_{ab}=
\begin{cases}
\frac{-\bar {\alpha}_{ab} \bar c_{ab}\omega_{ab} + \beta \omega_{ab}^2}{\bar \rho_{ab}}& \text{for~~}\omega_{ab}  < 0\,,\\
0 &  \text{for~~} \omega_{ab}\ge 0\,,
\end{cases}
\label{eq:AV_sphynx}
\end{equation}

 \noindent where $\omega_{ab}$ is given by Eq.~\ref{eq:omega} and $\alpha_{min}\le \alpha_a\le\alpha_{max}$, while $\beta$ remains constant. SPHYNX and SPH-EXA also employ Eq.~\ref{eq:AV_sphynx} to implement dissipation. Typical values of the AV coefficients when using switches are: $\alpha_{min}=0.05$, $\alpha_{max}=1$, and $\beta=2$.

\begin{table*}[t]
\centering
\caption{AV variants considered in this work.} 

\begin{tabular}{lccccl}
\toprule
Method & SLR & $\mathcal{B}$ modulation& Switches  & $\alpha$ coefficient & Notes \\
\midrule
1: AV         & No  & --                    & No   &  $\alpha=1$ & Classical shock-capturing baseline.\\
2: AVSW       & No  & --                    & Yes  & $\alpha_a\in[\alpha_{\min},\alpha_{\max}]$ & Time-dependent $\alpha$. \\
3: AVSLR      & Yes & No                    & No   & $\alpha=1$ & Slope-limited reconstruction.\\
4: AVSWSLR    & Yes & No                    & Yes  & $\alpha_a\in[\alpha_{\min},\alpha_{\max}]$& Hybrid: switches + SLR. \\
5: AVSLRB     & Yes & $(1-\mathcal{B})$ & No  & $\alpha=1$ & No SLR when $\mathcal B \simeq 1$; strong SLR when $\mathcal B\simeq 0$. \\
6: AVSLRB2    & Yes & $(1-\mathcal{B}^{\,2})$ &  No &  $\alpha=1$ & No SLR when $\mathcal B \simeq 1$; strong SLR if $\mathcal B\lesssim 0.3$. \\
7: AVSLRB2${\mathcal F}$ & Yes &  $(1-\mathcal{B}^{\,2})$  & No  & $\alpha = \alpha_{max}\times max[{\mathcal F}, \mathcal B]$ & Hybrid: SLR + $\alpha$ with floor value $\mathcal F$.\\
\bottomrule
\end{tabular}
\tablefoot{SLR denotes the slope-limited velocity reconstruction of Eqs.~(\ref{eq:filter_a})–(\ref{eq:filter_b}), optionally modulated by the Balsara factor $\mathcal B$ ($0\le \mathcal{B}\le 1$) via Eqs.~(\ref{eq:filter_c})–(\ref{eq:filter_d}). Switches refers to time-dependent $\alpha(\mathbf{r},t)$. The last row is discussed in relation to the subsonic turbulence scenario (Sec.~\ref{sec:turbulence}). The coefficient $\beta$ is always fixed to $\beta=2$.}
\label{tab:av_variants}
\end{table*}

\subsection{Artificial viscosity with shock detection: Switches}

The choice of $\alpha$ and $\beta$, which control the linear ($\propto\omega_{ab}$) and the quadratic ($\propto \omega_{ab}^2$) contributions to the AV in Eq.~\ref{eq:AV_sphynx}, is crucial in practical applications of SPH. The simplest option is to keep $\alpha~ \mathrm{and}~ \beta$ constant during the calculation. However, values of $\alpha$ large enough to capture strong shocks introduce excessive viscosity in smooth regions of the fluid. Overdissipation damps short-wavelength modes and hinders the growth of fluid instabilities and turbulence. It also suppresses differential rotation, resulting in a poor depiction of, for instance, the accretion disk dynamics. 

A plausible, albeit phenomenological, remedy was proposed by \cite{bal95}: adopting constant $\alpha=1$ and $\beta=2$, but multiplying $\Pi_{ab}$ by a limiter ${\mathcal B}_{ab}$ 
designed to approach unity in compressive (shock) regions and to decrease toward zero in curl-dominated flows,

\begin{equation}
\Pi_{ab}\xleftarrow{}{\mathcal B}_{ab}\Pi_{ab}\,;\qquad {\mathcal B}_{ab}=\frac{1}{2}({\mathcal B}_a+{\mathcal B}_b)\,,
\label{eq:balsara_0}
\end{equation}

\noindent with

\begin{equation}
{\mathcal B}_{a}= \frac{\vert\nabla\cdot \mathbf v\vert_a}{\vert\nabla\cdot \mathbf v\vert_a + \vert \nabla\times \mathbf v\vert_a+10^{-4}\frac{c_a}{h_a}}\,, 
\label{eq:balsara_1}
\end{equation}

\noindent where $c_a$ and $h_a$ are the speed of sound and the smoothing length of particle $a$, respectively. 

Later, a more efficient algorithm was introduced by \cite{morris1997} and subsequently refined by others. In this approach, the linear AV coefficient is kept near a low default value, $\alpha=\alpha_{min}$ (with $\alpha_{min}\simeq 0.05$), but quickly increases towards $\alpha_{max}$ when a shock is detected. \cite{cul10} showed that fast and robust shock detection is crucial in this technique and proposed using a higher order indicator based on the derivative of $\nabla\cdot{\bf v}_a$, which generally yields improved results. Several variations and refinements of these switches for controlling $\alpha ({\bf r}, t)$ can be found in the literature \citep{read2012, ros15}.  By default, SPHYNX and SPH-EXA use the same high-order dissipation switch proposed in \cite{read2012}, which can detect flow convergence before it occurs and can be considered the state-of-the-art version of that technique.

 The specific implementation is the following: $\alpha_a$ is initialized at a low value (usually $\alpha_a=0.05$). Then it increases rapidly if dissipation is needed and decays exponentially to its original value when it is not needed, effectively acting as a switch for the AV at the position of each SPH particle.

We start by calculating a local viscous parameter, $\alpha_{loc,a}$ \citep{read2012}, expressed as

\begin{equation}
\alpha_{loc,a}=
\begin{dcases}
\frac{\alpha_{max}\mathcal{D}_a}{\mathcal{D}_a+h_a\vert\nabla\cdot \mathbf{v}_a\vert+0.05 c_a} & \text{for~~$\nabla\cdot\mathbf{v}_a < 0$}\,,\\
 0 & \text{otherwise}\,,
\end{dcases}
\label{eq:alphaloc}
\end{equation}

\noindent
where $\mathcal{D}_a=h^2_a\vert\nabla(\nabla\cdot\mathbf{v}_a)\vert$ predicts flow convergence by comparing the gradient of $\nabla\cdot\mathbf{v}_a$ with the local value of $\nabla\cdot\mathbf{v}_a$. If the former is large enough and the fluid is actually converging ($\nabla\cdot\mathbf{v}_a<0$), $\alpha_{loc,a}$ adopts a value between 0 and $\alpha_{max}$. Furthermore, the term $0.05 c_a$, where $c_a$ stands for the local speed of sound, acts as a base value for the size of the velocity fluctuations that can trigger dissipation, and it also works as a term to prevent divergent $\alpha_{loc,a}$. Once we know $\alpha_{loc,a}$ we have two options:
\begin{itemize}
    \item if $\alpha_{loc,a}\geq\alpha_a$: we instantaneously raise the dissipation setting $\alpha_a=\alpha_{loc,a}$\,,
    \item if $\alpha_{loc,a}<\alpha_a$: we let $\alpha_a$ decay smoothly.\,
\end{itemize}

In the case that $\alpha_a$ has to decay, we evaluate an instantaneous decay rate of $\dot{\alpha}_a$ \citep{read2012}, expressed as
\begin{equation}
    \dot{\alpha}_a=
    \begin{dcases}
        (\alpha_{loc,a}-\alpha_a)/\tau_a & \text{if~~$\alpha_{min}<\alpha_{loc,a}<\alpha_a$},\\\
        (\alpha_{min}-\alpha_a)/\tau_a & \text{if~~$\alpha_{min}\geq\alpha_{loc,a}$}\,.
    \end{dcases}
    \label{eq:alphadot}
\end{equation}

The decay time, $\tau_a$, is defined as $\tau_a=h_a/v_{sig,max}$ and it is a function of the local spatial resolution and the maximum signal velocity, $v_{sig,max}$, in the neighborhood of particle $a$, calculated according to Eq.~\ref{eq:signalvel}. The effect of Eq.~\ref{eq:alphadot} is to allow alpha to decay exponentially to $\alpha_{loc,a}$ (or $\alpha_{min}$) if $\alpha_{loc,a}$ is too low. In this way, we can ensure a minimum dissipation that helps reduce the random noise that might arise from particle disorder. However, we note that even though all particles have at least a minimum value, $\alpha_{min}$, dissipation is automatically cut off when $\mathbf{r}_{ab}\cdot\mathbf{v}_{ab}\geq0$. Also, there is no restriction for $\alpha_{min}$, which can be set to zero.

At this point, it is worth highlighting the subtle difference between a limiter and a switch in this context. A limiter,
such as the  Balsara one, is a continuous, instantaneous, memoryless amplitude control that keeps dissipation high by default, except in regions that meet well-defined physical criteria (e.g., a high curl value). A switch, on the other hand, is a tunable sensor with temporary memory that can lag or overshoot in different regions. Switches are engineered to work well in some tests, but their exact form is a practical choice, rather than a unique consequence of the equations.

Despite their success, AVSWs can introduce additional noise, especially in the post-shock wake where $\alpha$ may oscillate due to rapid changes in the shock indicator, leading to spurious entropy production \citep[e.g.,][]{cul10,read2012,ros15}. In mixed regimes where shocks and strong shear coexist (e.g., when a dilute stream impacts a dense cloud), switches might struggle to balance compression-driven dissipation against shear suppression \citep[cf.][]{bal95,price18}. Finally, switches tend to be less effective in controlling low-level numerical noise in otherwise quiescent regions; dedicated “noise triggers” can be added to mitigate this, but at the expense of additional heuristics and computing time \citep{ros15}.

\begin{table*}
\centering
\small
\caption{Simulations name convention and some features of the studied test cases.}

\begin{tabular}{ccccccccccc}
\hline
\multirow{2}{*}{Test} & AV & AVSW  & AVSLR & AVSWSLR & AVSLRB & AVSLRB2 & AVSLRB2${\mathcal F}$ &Box& $N$ & \multirow{2}{*}{$N^{\mathrm{1D}}$}\\ 
 &$\alpha=1$ & $\alpha(t)$ & $\alpha=1$ & $\alpha (t)$ & $\alpha=1$ & $\alpha=1$ & $\alpha=1$, ${\mathcal F}=0.5$ &$[L_x,L_y,L_z]$ & $(\times 10^6)$&\\
\hline      
Shock-Tube & ST$_1$ & ST$_2$ & ST$_3$ & ST$_4$ & ST$_5$ & ST$_6$ & ST$_7$ & $[1,1/8,1/8]$ & 0.97 & 396\\
Sedov      & S$_1$  & S$_2$  & S$_3$  & S$_4$  & S$_5$  & S$_6$  & S$_7$  & $[1,1,1]$     & 8.0  & 200\\
GH-Vortex  & V$_1$  & V$_2$  & V$_3$  & V$_4$  & V$_5$  & V$_6$  & V$_7$  & $[1,1,0.5]$   & 0.5  & 100\\
KHI        & KH$_1$ & KH$_2$ & KH$_3$ & KH$_4$ & KH$_5$ & KH$_6$ & KH$_7$ & $[1,1,0.1]$   & 0.6  & 182\\
RTI        & RT$_1$ & RT$_2$ & RT$_3$ & RT$_4$ & RT$_5$ & RT$_6$ & RT$_7$ & $[1,2,0.1]$   & 1.2  & 182\\
Wind-Cloud & WC$_1$ & WC$_2$ & WC$_3$ & WC$_4$ & WC$_5$ & WC$_6$ & WC$_7$ & $[40,10,10]$  & 0.5  & 200\\
Turbulence & T$_1$  & T$_2$  & T$_3$  & T$_4$  & T$_5$  & T$_6$  & T$_7$  & $[1,1,1]$     & 512  & 800\\
\hline
\end{tabular}   
\tablefoot{The first column is the test type. The second to seventh columns show the test name that relates the test type to the methods defined in Table~\ref{tab:av_variants}. 
The three last columns indicate the size of the computational box, the total number of SPH particles, and the equivalent number per axis, N$^{\mathrm {1D}}$ in a cubic box with $L=1$, calculated as $N^{\mathrm {1D}}=\left[N/(L_x\cdot L_y\cdot L_z)\right]^{1/3}$. $\beta=2$ in all tests.}
\label{tab:table_cases}
\end{table*}

\subsection{A shock-capturing algorithm free of switches} 

A linear velocity field implies a smooth flow without the presence of shocks. Therefore, linear velocity fields should have a vanishing AV. Following the prescription introduced by \cite{fro17}, we reconstructed the local velocity field for each particle by removing the linear component of the velocity jump that is used to trigger and calculate the dissipation (Eq.~\ref{eq:AV_sphynx}). Assuming differentiability in the neighborhood of each SPH particle, the optimal linear approximation of the velocity field is given by its Jacobian acting on the separation vector. Rather than using $\vv_{ab}=\vv_a-\vv_b$ in the AV, we use $\vv'_{ab}=\vv'_a-\vv'_b$, defined as

\begin{align}
      \label{eq:filter_a}
    v'_{a,i}&\equiv v_{a,i}-\frac{1}{2}\phi_{ab}\mathbf{J}_{\vv_a}\vx^T_{ab}\,,\\
    v'_{b,i}&\equiv v_{b,i}+\frac{1}{2}\phi_{ba}\mathbf{J}_{\vv_b}\vx^T_{ab}\,,
    \label{eq:filter_b}
\end{align}

\noindent
where $\vx^T_{ab}=\vx_a-\vx_b$ is the column vector of the relative distance between particles $a$ and $b$, and $\mathbf{J}_{\vv}$ is the Jacobian of the velocity field. In the presence of discontinuities, such a linear estimator is not accurate enough; hence, the factor $\phi_{ab}$ is designed to approach zero in these cases and to remain close to unity for smooth flow. \citet{fro17} proposed a van Leer-like limiter \citep{vanLeer1974}:

\begin{align}
    \phi_{ab} &=\max\left[0,\min\left(1,\frac{4F_{ab}}{(1+F_{ab})^2}\right)\right]\times\kappa_{ab}\,,\label{eq:phiab}\\
    \kappa_{ab} &\equiv
    \begin{dcases}
        \exp{\left[-\left(\frac{q_{ab}-q_{crit}}{q_{fold}}\right)^2\right]} & \text{if~~$q_{ab}<q_{crit}$}\,,\\
        1 & \text{if~~$q_{ab}\geq q_{crit}$}\,,
    \end{dcases}\label{eq:kappaab}\\
    q_{ab} &\equiv \min(q_a,q_b)\,,\\
    q_{crit} &\equiv \left(\frac{32\pi}{3nb_a}\right)^{1/3}\,,\\
    F_{ab} &\equiv \frac{\vx_{ab}\mathbf{J}_{\vv_a}\vx^T_{ab}}{\vx_{ab}\mathbf{J}_{\vv_b}\vx^T_{ab}}\,,
\end{align}

\noindent
where $q_{a}=|\vx_b-\vx_a|/h_a$, $q_{crit}$ estimates the mean interparticle spacing in units of smoothing length, while $nb_a$ is the neighbor count of particle $a$. The parameter $q_{fold}$ sets the width of the Gaussian correction (we adopted the recommended value from \cite{fro17} of $q_{fold}=0.2$). 

Equations \ref{eq:filter_a} and \ref{eq:filter_b} remove the linear component of the velocity field, limiting the slope at discontinuities, while reducing unwanted dissipation in smooth flows. We refer to this slope-limited reconstruction as AVSLR, and contrast it with  our implemented AVSW (time-dependent switch) method in what follows.  We note that although both AVSLR and AVSW share a heuristic origin, linear reconstruction does not involve time derivatives and decay laws.

Currently, few calculations incorporating these AVSLRs have been reported \citep{fro17,rosswog2020,rosswog2020magma, cabezon2025, sandnes2025}. They were mostly used in combination with switches. As in \cite{price2019} and \cite{sandnes2025}, our premise is that AV switches are unnecessary when using the slope-limited reconstruction algorithm. We therefore explored several variants of the implementation of the SLR (see Table~\ref{tab:av_variants}), as follows:

 \begin{itemize}
 
     \item  AVSLR: use Eqs.~\ref{eq:filter_a}, \ref{eq:filter_b} to remove the linear component of the velocity field;

     \item   AVSLRB: modulate the SLR  with the Balsara limiter:

     \begin{align}
        v'_{a,i}&\equiv v_{a,i}-\frac{1}{2}\left(1-{\mathcal B}^p_a\right)\phi_{ab}\mathbf{J}_{\vv_a}\vx^T_{ab}\,,\label{eq:filter_c}\\
        v'_{b,i}&\equiv v_{b,i}+\frac{1}{2}\left(1-{\mathcal B}^p_b\right)\phi_{ba}\mathbf{J}_{\vv_b}\vx^T_{ab}\,,\label{eq:filter_d}
     \end{align}

\noindent with ${\mathcal B_{a,b}}$ defined by Eq.~\ref{eq:balsara_1}\footnote{In the tests performed here, we used the average  $\mathcal {B}_{ab} = \frac{1}{2}(\mathcal{B}_a+\mathcal{B}_b)$ in Eqs.~\ref{eq:filter_c} and \ref{eq:filter_d} so that the multiplicative factor  $\frac{1}{2}(1-\mathcal{B}_{ab}^p)$ is the same in both expressions. }. Following the suggestion of \cite{sandnes2025}, we first consider the case with $p=1$.

     \item AVSLRB2 same as above but with $p=2$; this does not change the description of shocks, although it proved more advantageous than $p=1$ in our tests.
     
     \item  Baselines: for comparison, we also consider (i) classical dissipation with fixed coefficients (AV); (ii) AV with time-dependent switches (AVSW); and (iii) a hybrid combining switches with slope-limiting (AVSWSLR).
The objective is to find the most balanced and general scheme among all
these options in light of the results.       
  \end{itemize}

\section{Tests}

\label{sec:tests}

We first considered six 3D test problems designed to compare the performance of the proposed schemes across shock- and shear-dominated regimes. The shock-dominated set comprises the Sod shock tube \citep{sod1978} and the Sedov blast \citep{sedov1959}. The shear and instability set comprises the steady Gresho–Chan vortex \citep{gresho1990}, Kelvin–Helmholtz (KH), and Rayleigh-Taylor (RT) instabilities, as well as the wind–cloud interaction where a dilute supersonic wind disrupts a dense cloud \citep{age07}. A seventh test dealing with subsonic turbulence is deferred to Sect.~\ref{sec:turbulence}. When needed, we adopted an ideal-gas equation of state $P=(\gamma-1)\rho u$, with $\gamma=5/3$ (or with $\gamma=1.4$ and $\gamma=1.001$ for the RT and turbulence tests, respectively). Simulations were carried out with SPHYNX\footnote{\url{https://astro.physik.unibas.ch/sphynx}} \citep{cabezon2017}, with the exception of turbulence, which was calculated with SPH-EXA\footnote{\url{https://github.com/sphexa-org/sphexa}.}, including the updates discussed in \citealt{garciasenz2022}). Kernel interpolations use the mixed sinc kernel $S^{0.9}_{4,9}$ \citep{cabezon_mixedsincs_2024}, which is pairing-resistant and robust for a wide range of neighbors, $60\le n_b \le 400$. Unless stated otherwise, we set $n_b=100$. Additional setup details for each case are summarized in Tables~\ref{tab:av_variants} and \ref{tab:table_cases}.

\begin{figure}[h]
\centering
\includegraphics[width=\columnwidth]{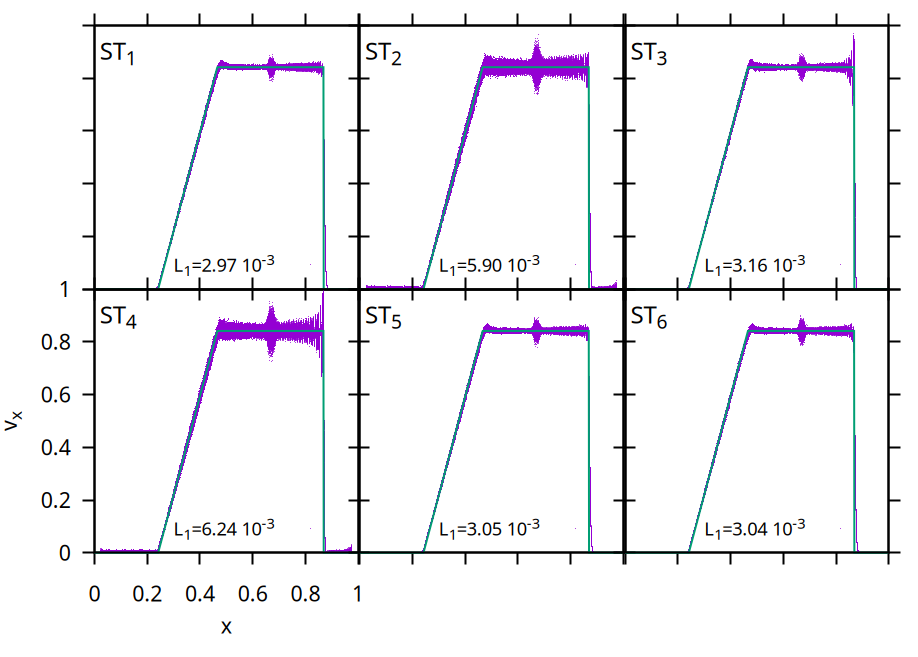}
\caption{Velocity profile, $v_x$, for the shock-tube test and the models shown in the second row in Table \ref{tab:table_cases}, at the time $t=0.2$}
    \label{fig:shock-tube}
\end{figure}

\begin{figure}
\centering
\includegraphics[width=0.5\textwidth]{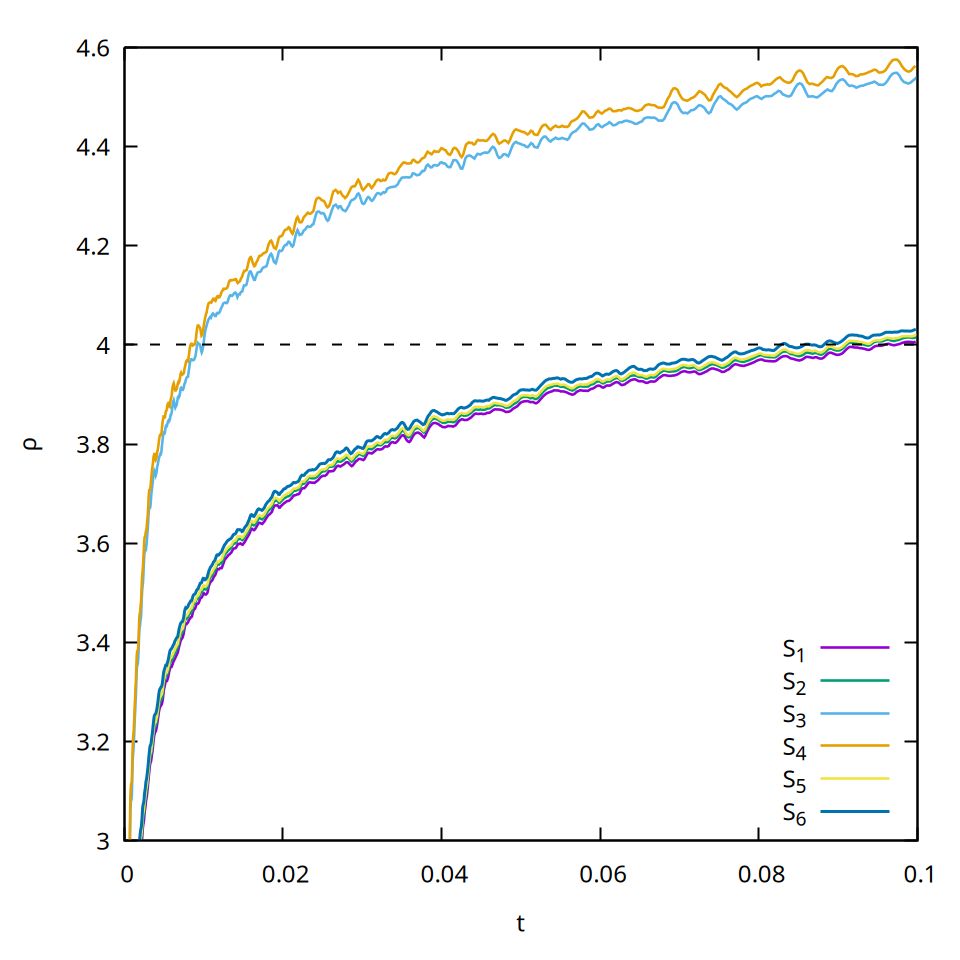}
    \caption{ Maximum density evolution in the Sedov test for the range of models in Table \ref{tab:table_cases}.}
    \label{fig:sedov_0}
\end{figure}
To make quantitative comparisons between models, we report a mean absolute error $L_1$ wherever a suitable reference is available,
\begin{equation}
   L_1 = \frac{1}{N_{\mathcal S}} \sum_{i\in \mathcal S} \bigl|p^{\rm SPH}_i - p^{\rm ref}_i\bigr|\,,
\label{eq:L1}
\end{equation}

\noindent 
where $S$ denotes the set of sampling points with positions $d_i$ in the range $d_0\leq d_i\leq d_f$ (e.g., a 1D line-out or radial profile), and $N_S$ is the number of particles belonging to the set, $S$. Here, $p^{\rm SPH}$ is the measured quantity from the simulation and $p^{\rm ref}$ is the corresponding reference (analytic solution or high-resolution benchmark) specified for each test below.

\subsection{The shock-tube}
\label{subsec:shock-tube}

In this test, a box with the size and number of equal-mass particles specified in Table~\ref{tab:table_cases} is initialized with a gas that has a strong discontinuity in density and pressure across a contact interface at $(x=x_c)$. The problem has a well-known analytical solution \citep{sod1978}, which we use to assess the performance of the different shock-capturing prescriptions.

\begin{figure}
\centering
\includegraphics[width=\columnwidth]{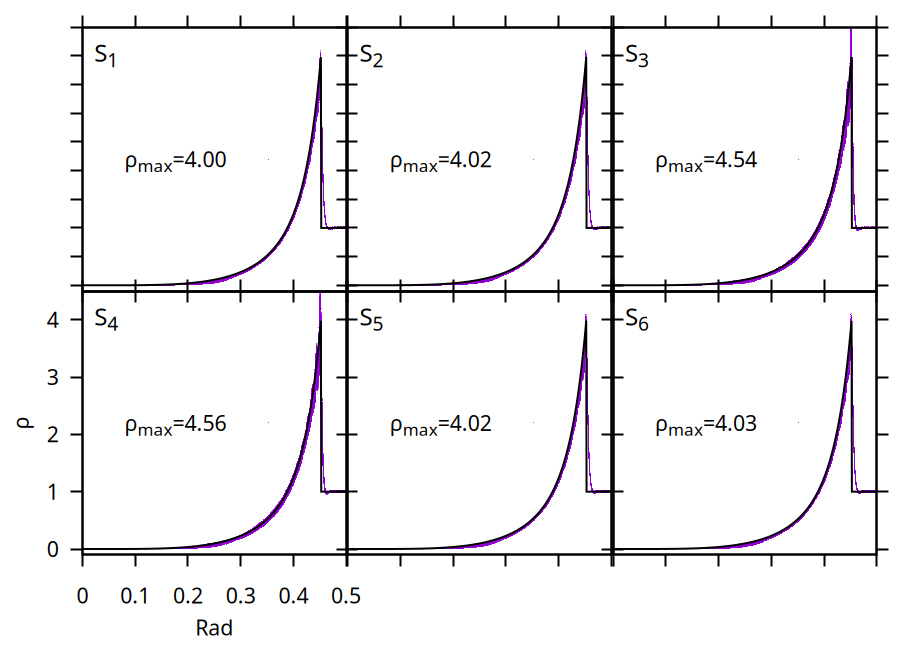}
\includegraphics[width=\columnwidth]{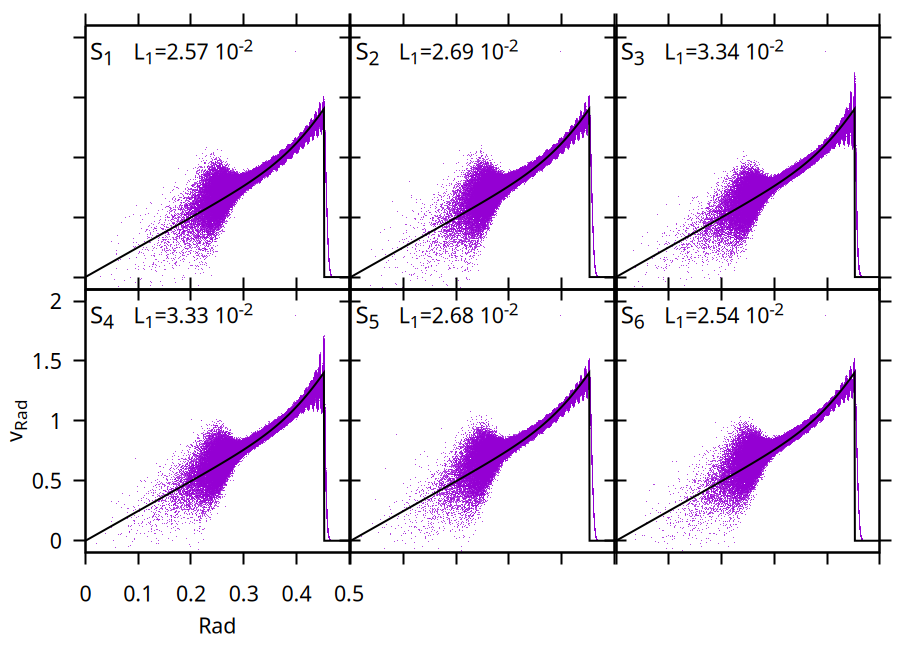}
    \caption{Density and radial velocity profiles in the Sedov-Taylor test at time $t=1$ for the different models described in Table~\ref{tab:table_cases}. The solid line represents the analytical reference. The $L_1$ error between the numerical and analytical estimates from the center until the maximum radial velocity is also shown.}
    \label{fig:sedov}
\end{figure}

The setup follows common practice \citep{Wadsley17, price18}: in the left state $(x< x_c)$, we take $\rho_l=1$, $p_l=1$; in the right state $(x> x_c)$, $\rho_r=0.125$, $p_r=0.10$. The particle distribution across the contact is left unsmoothed, but the pressure was smoothed according to

\begin{equation}
 P_0= -A~\tanh \left(\frac{x-x_c}{\delta}\right)+B\,,
\end{equation}

\noindent with $\delta=0.01$, $x_c=0.5$, $A=0.45$, and $B=0.55$. 

This test probes how the different AV choices handle strong shocks. The results are summarized in Fig.~\ref{fig:shock-tube}, which shows the $x$-velocity profile, $v_x$. We computed the $L_1$ error using Eq.~\ref{eq:L1} over $x\in[x_0,x_f]$ with $x_0=d_0=0.05$ and $x_f=d_f=0.868$. The upper limit of $x_f=0.868$ was chosen to exclude the terminal vertical drop in the velocity, so that the sum focuses on the resolved shock, contact, and rarefaction features.

From Fig.~\ref{fig:shock-tube}, models ST$_2$ and ST$_4$ (i.e., those employing switches) show the worst performance, exhibiting pronounced velocity oscillations in the post-shock region. Substantially better results are obtained either with the simple, constant-parameter AV (ST$_1$) or with the SLR approach, modulated with the Balsara limiter (ST$_5$ and ST$_6$). The ST$_3$ calculation is intermediate, with oscillations confined to the immediate shock vicinity. Overall, the best performance in this set is delivered by ST$_1$, followed by ST$_6$, which uses SLR and applies the squared ($p=2$) Balsara limiter to the functions in Eqs.~\ref{eq:filter_c} and \ref{eq:filter_d}.

\subsection{Spherical explosion}

\begin{figure}
\centering
\includegraphics[width=\columnwidth]{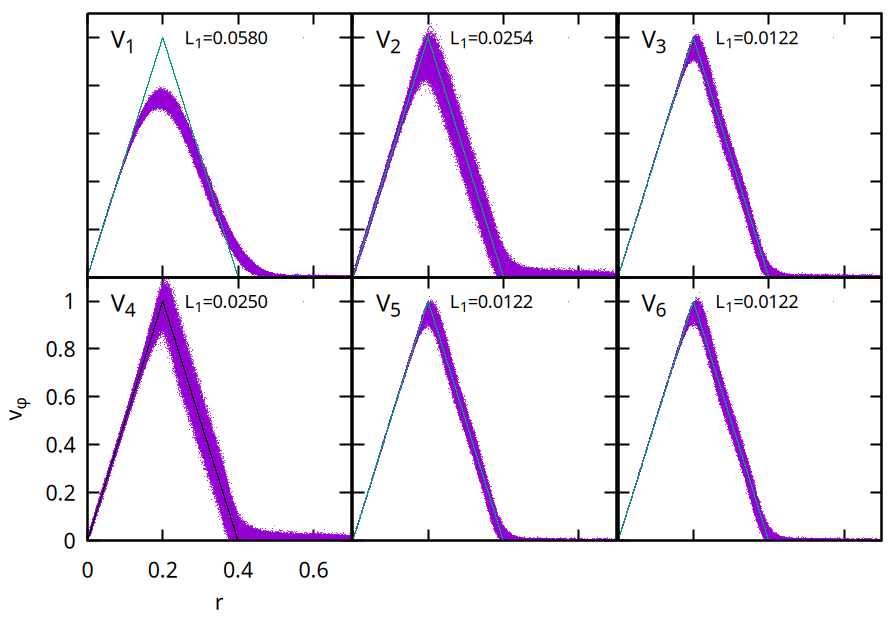}
    \caption{Azimuthal velocity distribution in the Gresho-Chan vortex experiment at $t=1$ for  the model V$_k$ in the third row in Table \ref{tab:table_cases}. The analytical reference is indicated by the continuum line.} 
    \label{fig:vortex1}
\end{figure}

Adopting the test that is typically referred to as the Sedov-Taylor blastwave, we performed it in a 3D box with a side $L=1$ with $N=200^3$ particles arranged in a glass-like configuration. All particles have the same mass, resulting in a homogeneous initial density profile of $\rho=1\pm 0.005$. The explosion is initiated in the center of the box by depositing a total energy $\Delta U=1$ into the gas, distributed with a Gaussian profile of characteristic width $\delta=2 h$, where $h$ is the smoothing length at $t=0$.

Figures \ref{fig:sedov_0} and \ref{fig:sedov} summarize the results. The time evolution of the maximum density is shown in Fig.~\ref{fig:sedov_0}. According to these figures, models S$_3$ and S$_4$ overshoot the strong-shock compression limit $\rho/\rho_0=(\gamma+1)/(\gamma-1)=4$ for $\gamma=5/3$, while the remaining four models stay below and approach this limit from below as the blast develops. As in the shock-tube test, we computed the $L_1$ error. In this case, for the radial velocity between the origin and the r coordinate of the maximum velocity. The lower panels of Fig.~\ref{fig:sedov} show the profile of $v_{rad}$ and the $L_1$ value at $t=0.1$. In the central region ($r\le 0.25$), the numerical profiles depart from the analytical solution (solid line), although the amount of mass involved is very small. To adequately resolve the diluted central region, a much larger number of SPH particles would be needed. In the post-shock region, all the models exhibit velocity oscillations, which are particularly pronounced near the shock front for models S$_3$ and S$_4$, consistent with the behavior observed in the shock-tube test.

According to these results, the best models are S$_6$, S$_1$, and S$_5$ (these three have very similar results), followed by S$_2$. The worst results arise when using switches together with the SLR technique (S$_4$) or with SLR alone without Balsara's modulation (S$_3$).

\subsection{The Gresho-Chan vortex}

The simulation of a pure fluid vortex \citep{gresho1990} is challenging for SPH schemes working with standard AV: excessive viscosity damps the differential rotation and degrades the velocity field. The flow is axisymmetric and defined by

\begin{equation}
  v_\varphi(r)=v_0
\begin{cases}
    \psi & \text{for $\psi\leq 1$}\,,\\
    2-\psi & \text{for $1<\psi\leq 2$}\,,\\
    0 & \text{for $\psi>2$}\,,
\end{cases}
\label{eq:vortex_1}
\end{equation}

\begin{equation}
  P(r)= P_0+4v_0^2
\begin{cases}
    \psi^2/8 &  \text{for $\psi\leq 1$}\,,\\
    \left(\psi^2/8-\psi +\ln \psi +1\right) &   \text{for $1<\psi \leq 2$}\,,\\
    \left(\ln 2 -1/2\right) & \text{for $\psi>2$}\,,\\
\end{cases}
\end{equation}

\noindent with $\psi=5r$, $r=\sqrt{x^2+y^2}$, $v_0=1$, $\rho_0=1$, and $P_0=5$.

The box size and number of SPH particles are listed in Table~\ref{tab:table_cases}. Because the initial azimuthal velocity is piecewise linear, this test is expected to favor AVSLRs methods over AVSWs. The results are summarized in Fig.~\ref{fig:vortex1}, which also provides the $L_1$ error for quantitative comparisons. We compute $L_1$ using Eq.~\ref{eq:L1} over $r\in[d_0,d_f]=[0,0.5]$, with $p^{SPH}, p^{ref}$ taken as the azimuthal velocity from the SPH calculation and the analytic $v_{\varphi}$ in Eq.~\ref{eq:vortex_1}, respectively.

Figure~\ref{fig:vortex1} clearly shows the superior performance of AVSLRs schemes over AVSWs. As expected, the worst case is V$_1$, which uses AV with constant coefficients $\alpha=1$, $\beta=2$, and no limiter. Introducing switches yields a substantial improvement, but still falls short of the SLR methods. For this test, the best results are obtained with V$_4$ (AVSRL), V$_5$ (AVSLRB), and V$_6$ (AVSLRB2), with practically no differences between them.

\begin{figure}[h]
\centering
\includegraphics[width=\columnwidth]{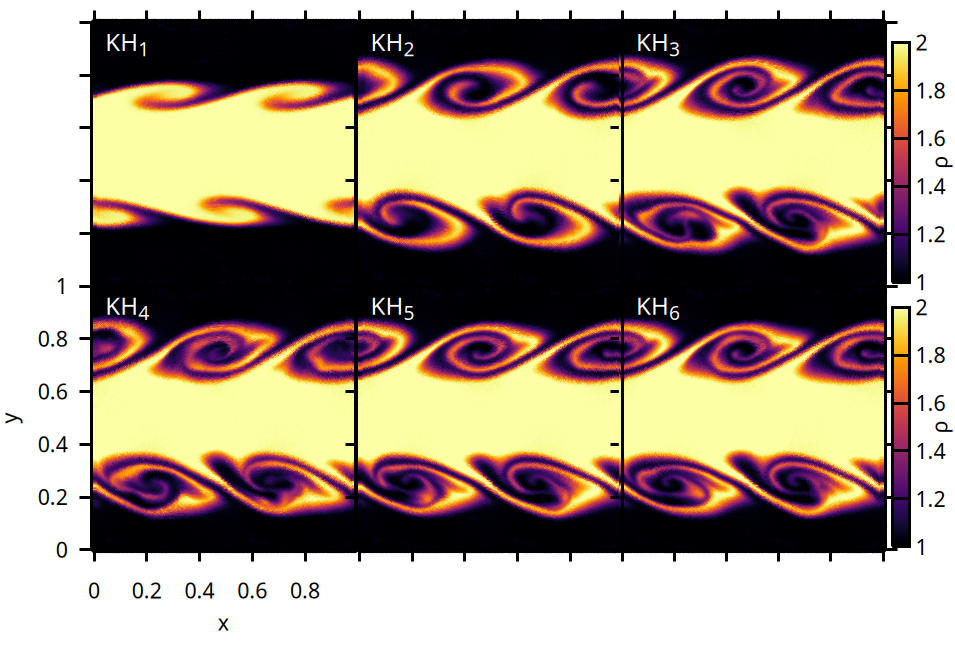}
\caption{Color-coded density for the KH models in Table~\ref{tab:table_cases} at $t=3$ ($t/t_{\mathrm {KH}}=2.8$).}
\label{fig:KH_1}
\end{figure}
    
\begin{figure}
\centering
\includegraphics[width=0.46\textwidth]{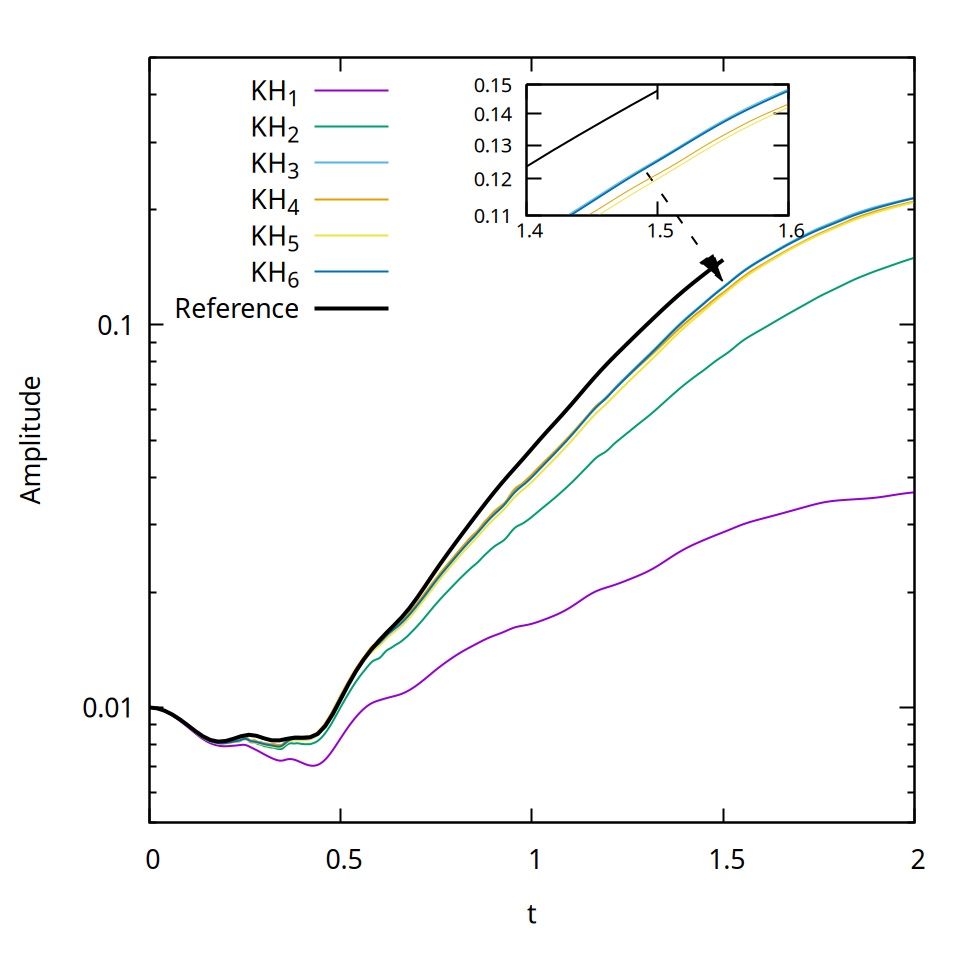}

    \caption{Amplitude mode growth of the KH models. A zoom-in of the nearby lines KH$_3$, KH$_4$, KH$_5$, and KH$_6$ around $t=1.5$ is shown in the subfigure.}
    \label{fig:KH_2}
\end{figure}
 
\subsection{The Kelvin-Helmholtz instability}

The evolution of a shear layer between two fluids of different densities naturally triggers the Kelvin-Helmholtz instability (KHI). Because this instability is ubiquitous in nature, including astrophysical phenomena, its accurate simulation is of great importance. Since AV can artificially dampen KHI growth, this problem is a sensitive test for the SLR algorithm. 

We initialize SPH particles in a thin 3D box of the size listed in Table~\ref{tab:table_cases}, with a 2:1 particle number ratio between the high- and low-density regions. The initial setting consists of three stratified layers, with the central layer having the higher density.

\begin{figure}[h]
\centering
\includegraphics[width=0.495\textwidth]{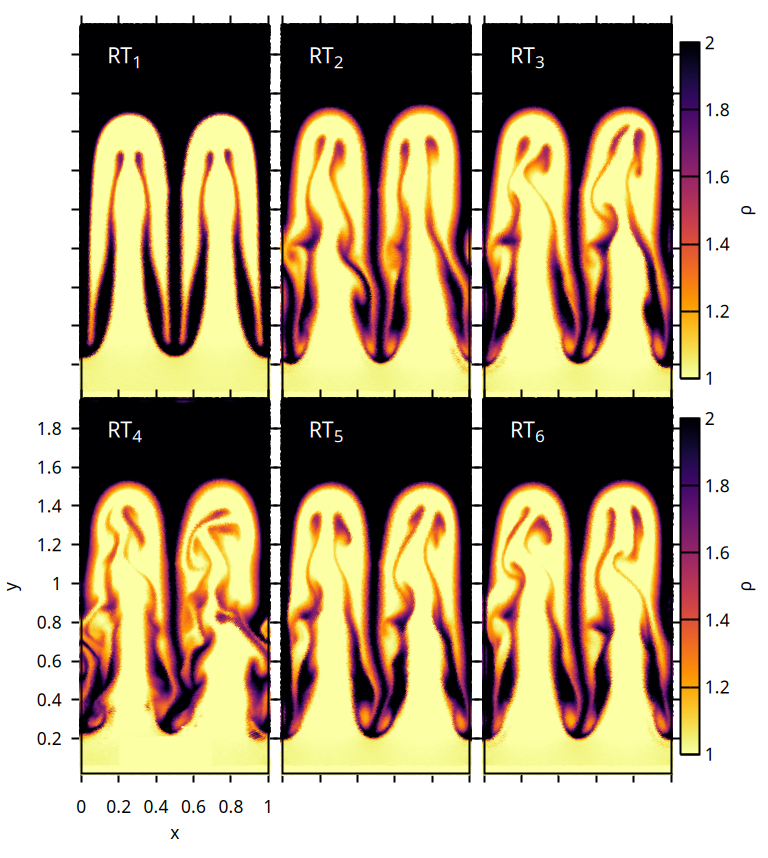}
 \caption{Color maps of density for the RT models in Table~\ref{tab:table_cases} at $t=6$.}
\label{fig:RT_1}
\end{figure}

The initial number of neighbors is $n_b=120$. The initial velocity distribution follows \cite{mcn12}, where shear was induced by imposing $v_x=\pm 0.5$ across the separation layer, exponentially damped along the $Y-$direction over a characteristic length, $L=0.025$. The density across the contact layer is not smoothed. We seed a single-mode normal perturbation in the cross-stream velocity, 

 \begin{equation}
     v_y(t=0) = \Delta v_y \sin\left(\frac{2\pi}{\lambda} x\right),\,
     \label{eq:kh_1}
 \end{equation}

 \noindent where $\Delta v_y=0.01; \lambda=0.5$ and we adopted an initial background pressure of \mbox{$P_0=2.5$}.

Times are also reported normalized to the characteristic KHI growth time $t_{\mathrm {KH}}$, as defined in \citet{age07}, resulting in $t_{\mathrm {KH}}=1.06$ for our models. The results are shown in Figs.~\ref{fig:KH_1} and \ref{fig:KH_2}, and Table~\ref{tab:table_KH}. Figure~\ref{fig:KH_1} shows the density color maps of the different models at a time, $t=3$ ($t/t_{\mathrm {KH}}=2.8$). At that time, the instability is fully nonlinear, with well-developed swirls, billows, and secondary structures. These extend over a wide range of wavelengths, with a cut-off set by the real viscosity but, in practice, by the resolution and numerical dissipation.

According to Fig.~\ref{fig:KH_1}, models KH$_3$ to KH$_6$ develop secondary structure more robustly than KH$_1$ (AV with constant coefficients and no switches). Steering dissipation with switches (KH$_2$) is the second-worst. This is quantified in Fig.~\ref{fig:KH_2}, which displays the evolution of the perturbation amplitude. Then, KH$_3$ and KH$_6$ show the fastest growth. At $t=1.5$, the amplitudes (presented in Table~\ref{tab:table_KH}) are close to the reference \citep{mcn12}, which is remarkable given the moderate resolution in these simulations $N^{1D} = 182$.

We also see that the use of switches (KH$_2$), represents a considerable improvement over AV with constant coefficients (KH$_1$). Even so, the computed growth rate is much lower than that obtained with the SLR family, especially KH$_3$ and KH$_6$. The KH$_5$ and KH$_4$ models show slightly less amplitude growth than KH$_3$, but still larger than KH$_2$, indicating that they can be considered good options when we are looking to balance accuracy and robustness.
\begin{table} [h]
        \centering
        \caption{KH: Perturbation amplitude $A(t)$ at $t=1.5$ for the models in Table~\ref{tab:table_cases}.}
\begin{tabular}{ccccccc} 
                \hline
                - & KH$_1$ & KH$_2$  & KH$_3$ & KH$_4$ & KH$_5$ & KH$_6$ \\ 
                \hline      
$A\ (\times 10^{-2})$& 2.87& 8.31 & 12.58 & 12.14 & 12.01 & 12.52 \\
                \hline
        \end{tabular}  
        \tablefoot {The reference value from \citet{mcn12} is $A(t=1.5)= 14.79\times 10^{-2}$}
        \label{tab:table_KH}
\end{table}

\subsection{The Rayleigh-Taylor instability}

The study of convective-like instabilities in a gravitational field is of major importance in astrophysics. Historically, SPH has struggled to model these systems. In particular, the growth of the Rayleigh-Taylor instability (RTI) was often poorly depicted because AV introduces excessive damping and kills the smallest perturbation modes. However, combining improved gradient estimators with AVSW made a reasonable approach to this phenomenon possible \citep{cabezon2017, rosswog2020}.

\begin{table}
    \centering
    \caption{RTI: Kinetic energy (KE) in the Y-direction at $t=6$ for the models in Table~\ref{tab:table_cases}.}
    \begin{tabular}{ccccccc} 
        \hline
        - & RT$_1$ & RT$_2$  & RT$_3$ & RT$_4$ & RT$_5$ & RT$_6$ \\ 
        \hline      
        KE ($\times 10^{-3}$)&6.81 & 8.49 & 8.87 & 8.39 & 8.62 & 8.87 \\
        \hline
    \end{tabular}   
    \label{tab:table_RT}
\end{table}

In this test, we used a box of size $[1, 2, 0.1]$ filled with a two-density fluid. The upper half has $\rho_u = 2$, whereas the lower half $\rho_d=1$. The system is immersed in a constant gravitational field with $g=-0.5$, and the initial pressure profile is set so that the hydrostatic balance is satisfied. The interface between the low- and high-density regions is sharp (no smoothing). The pressure at the bottom boundary is $P(y=0)=2.5$. The pressure profile is

\begin{equation}
  P\left(y\right)=
\begin{cases} 
    P\left(y=0\right)+\rho_d g y &  y \leq y_b -\frac{\delta}{2}\,,\\
     P\left(y_b-\frac{\delta}{2}\right) - \bar\rho g \left(y_b-\frac{\delta}{2}\right)+\bar\rho g y &   y_b-\frac{\delta}{2} < y < y_b + \frac{\delta}{2}\,,\\
    P\left(y_b+\frac{\delta}{2}\right)+ \rho_u g \left(y-\left(y_b+\frac{\delta}{2}\right)\right) & y \geq y_b+\frac{\delta}{2}\,,
\end{cases}
\end{equation}

\noindent where $y_b$ is the interface location, $\bar\rho$ is the average density around
$y_b$, and $\delta = 0.4$ is the initial width of the unstable layer $[y_b-\delta/2$, $y_b+\delta/2$]. The fluid is initially at rest, and we seed a velocity perturbation, $\Delta v_y$, localized to the interface, 

\begin{equation}
v_y=\Delta v_y \left[1-\cos\left(2\pi x\right)\right]\left[1+\cos\left(\frac{2\pi (y-y_b)}{\delta}\right)\right]\,,\  \mathrm{for}\ \vert y-y_b\vert \le \frac{\delta}{2}\,,
\label{eq:RTvel}
\end{equation}

\noindent with $\Delta v_y=0.025$. 

Despite the moderate resolution $N^{\mathrm{1D}}= 182$, we can assess the performance by inspecting the density maps in the nonlinear regime and the kinetic energy in the vertical direction. Figure~\ref{fig:RT_1} shows the color-coded density at $t=6$, while Table~\ref{tab:table_RT} provides quantitative information on the kinetic energy stored in vertical displacements at the same time. The RTI grows and becomes nonlinear even when with AV using constant $\alpha=1$, $\beta=2$ (RT$_1$), but without any discernible substructure. Introducing a time-dependent $\alpha$ (RT$_2$) significantly improves the outcome, enabling the development of secondary structures and larger bubble and spike amplitudes. Activating the SLR (RT$_3$) produces a similar qualitative improvement. However, although the AVSW calculation develops structure comparable to the AVSLR models, the stored kinetic energy in the vertical direction is clearly lower (Table~\ref{tab:table_RT}). Among the SLR variants, RT$_3$, RT$_5$, and RT$_6$ are similar, with RT$_3$ and RT$_6$ exhibiting the highest kinetic energy, and thus preferred, closely followed by RT$_5$.

Model RT$_4$, which combines switches and SLR, reduces AV via both mechanisms. The resulting lower dissipation amplifies numerical noise, leading to more asymmetric and more diffuse substructures. Despite the reduced viscosity, its vertical kinetic energy is the second-lowest in the set, reflecting the detrimental impact of noise on coherent RT growth.

\subsection{The wind-cloud test}
\label{WC}

The destruction of a dense cloud embedded in a dilute supersonic wind is demanding test that requires an accurate treatment of both shocks and shear-driven instabilities \citep[][and references therein]{age07}. That work showed that classic SPH underperformed relative to mesh methods, largely due to tensile instability that permeated SPH codes at the time. Modern SPH schemes, however, can handle this scenario well and yield competitive results \citep{fro17, Wadsley17, garciasenz2022}. While the mitigation of E0-errors for this problem  is more decisive than the choice of AV, the AV prescription still modulates the outcome and merits discussion.

\begin{figure}
\centering
\includegraphics[width=0.49\textwidth]{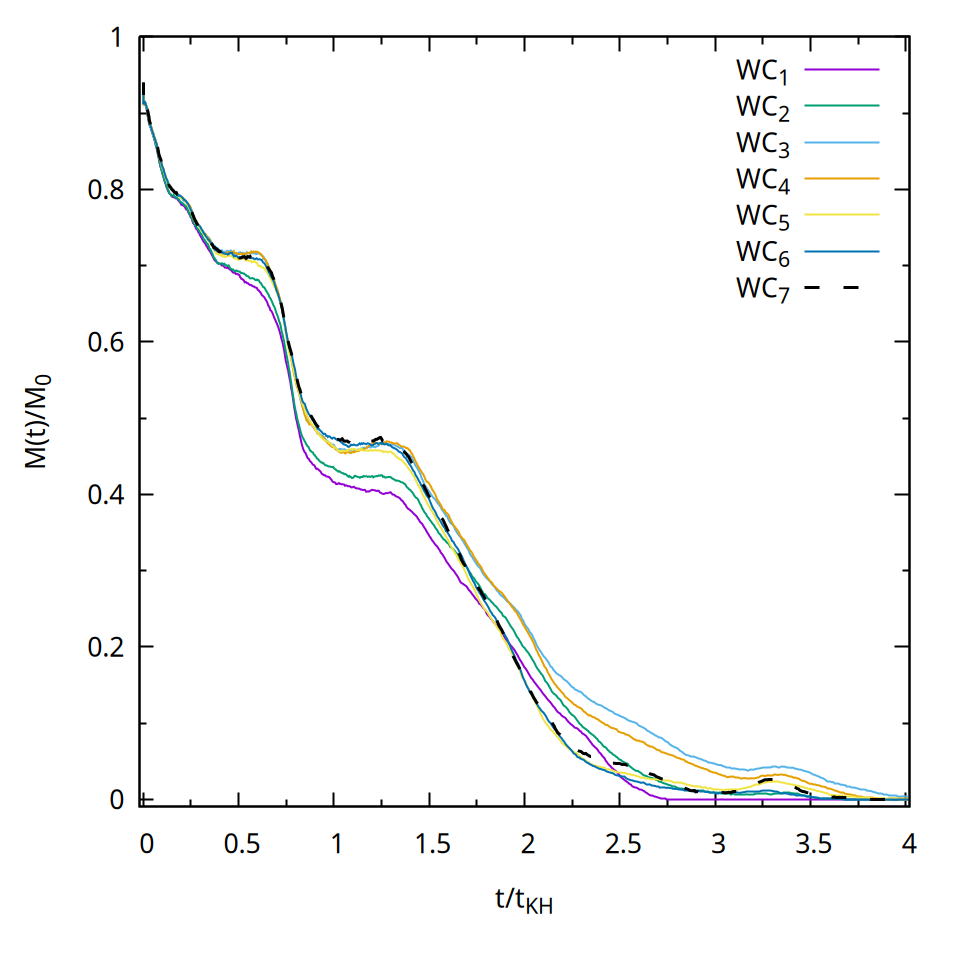}
    \caption{Evolution of the cloud mass normalized to its initial value. The elapsed time is normalized to $t_{KH}=2.9$.}
    \label{fig:WC_1}
\end{figure}

The initial configuration follows \cite{fro17}. The box size and particle count are given in Table~\ref{tab:table_cases}. The wind and cloud have initial densities of $\rho=1$ and 10, respectively, and are in pressure equilibrium at $P_0=1$. The cloud is initially at rest, while the wind moves supersonically with Mach number $\mathcal M=2.7$. Wind and cloud particles are distributed in glass-like lattices, and the interface is sharp. Numerical studies have identified different hydrodynamic regimes as a function of the KHI characteristic time $t_{KH}=2.9$ (coresponding to our setup\footnote{We take the characteristic $t_{KH}$ time as in \cite{age07}, which is defined as $t_{KH}=1.6~t_{crush}$, where $t_{crush}$ is the cloud crushing time, right after the formation of the bow-shock; beyond this, KHI  erodes the cloud surface.}) at which the KHI dominates:

\begin{itemize}
    \item $t\lesssim t_{KH}$.  Direct impact and bow-shock formation control the initial stripping of the material from the sphere. According to Fig.~\ref{fig:WC_1}, the cloud evolution is largely insensitive to the AV scheme.
    
    \item $t_{KH}\lesssim t \lesssim 3 t_{KH}$. Vorticity (primarily from KHI) peels and mixes cloud material into the wind. As seen in Fig~\ref{fig:WC_1}, the removal occurs at a remarkably steady rate. The destruction rate $\mathcal R = \dot M(t)/M_0$ differs across AV schemes and provides a convenient ranking criterion (Table~\ref{tab:table_WC}). 

    \item $t\ge 3t_{KH}$. A complete dissolution of the cloud is challenging to model physically and is also resolution-dependent, since mass remnants approach  zero at a sufficiently high resolution. In SPH, a small artificial thermal conductivity is commonly added to achieve complete mixing and thermodynamic equilibrium \citep{price08}. By default, SPHYNX includes an artificial heat conduction term, following \citet{price18} with a diffusion coefficient $\alpha_u=0.1$.
\end{itemize}

Overall, all AV schemes aim for complete cloud destruction at $t\le 4 t_{KH}$, in reasonable agreement with the results obtained with state-of-art mesh-codes such as ENZO  \citep[cited in][]{age07}. This reflects the strengths of our SPH flavor, namely, accurate gradient estimation via the integral approach \citep{garciasenz2012} and the geometric density average force estimation \citep{Wadsley17, garciasenz2022}. Still, the particular implementation of the AV is also important as it modulates the path to complete dissolution. The interval $1.25~t_{KH}\lesssim t\lesssim 2.5~t_{KH}$ is particularly relevant to our study because the slope of the $M(t)/M_0$ curves allows an easy classification of the AV variants.

According to Table~\ref{tab:table_WC}, the highest destruction rates at $1.25~t_{KH}\le t\le 2.5~ t_{KH}$ occur for WC$_6$, while the slowest are those with bare AV or with switches alone. Using this criterion, the AV schemes are ranked, from worst to best, in the same order as they appear in the table (we use the first row for ranking). However, note that SLR models without Balsara modulation (models WC$_3$ and WC$_4$) take longer to dissolve the cloud because they overestimate the effect of shocks, as seen in the Sedov test (Fig.~\ref{fig:sedov_0}); in the wind-cloud problem, this will delay efficient KHI-driven shredding.

\begin{table}
        \centering
        \caption{Cloud destruction rate, $\mathcal R= \dot M/M_0$, at $t=1.8\ t_{KH}$ and $t=2.0\ t_{KH}$ for the WC models.}
\begin{tabular}{ccccccc} 
                \hline
                 & WC$_1$ & WC$_2$  & WC$_3$ & WC$_4$ & WC$_5$ & WC$_6$ \\ 
                \hline      
$\mathcal R_{1.8t_{KH}}$&-0.105 & -0.112 & -0.139 & -0.144 & -0.152 & -0.162 \\
$\mathcal R_{2.0t_{KH}}$&-0.121 & -0.119 & -0.127 & -0.152 & -0.174 & -0.214 \\
                \hline
        \end{tabular}   
        \label{tab:table_WC}
\end{table}

\section{Subsonic turbulence}
\label{sec:turbulence}
Turbulence is ubiquitous across physical scales and is especially relevant in astrophysics, where it plays a major role in chemical mixing, stellar collapse, and supernova explosions, among other phenomena. High-resolution simulations of turbulence must be approached with care, as they are long and computationally expensive, with an adequate level of accuracy that must be maintained over many time steps. Subsonic turbulence is particularly demanding: even small accumulations of numerical error or a modest excess of dissipation can suppress the development of a faithful inertial range (i.e., the Kolmogorov cascade down to the dissipation scale). This has long challenged SPH approaches, which historically lagged behind Eulerian methods at low Mach numbers until recently. 

Since the Reynolds number in SPH with standard AV scales roughly as $\mathrm{Re} \propto \mathcal{M}$ at fixed resolution \cite{price12}, minimizing residual AV in smooth subsonic regions is essential to sustain an inertial range. Modern SPH implementations that combine accurate gradient estimators, generalized volume elements, pairing-resistant kernels, and tighter dissipation control have now achieved Eulerian-quality results \citep{cabezon2025}.

The simulation of subsonic turbulence presents an inverted dissipation problem relative to the previous tests in this work. In shocks, disks, and hydrodynamical instabilities, dissipation is driven primarily by strong compression and large velocity gradients. In contrast, subsonic turbulence is limited by the background numerical dissipation floor: if this baseline dissipation is even slightly too high, the energy cascade is prematurely damped, and the inertial range collapses. To mitigate spurious triggering by small-scale particle noise, it is therefore necessary to control the amplitude of the dissipation. To this end, we reuse the Balsara coefficients (already employed to modulate the SLR terms) to multiply the linear part of the AV, clamped from below by a floor, $\mathcal F$,

\begin{equation}
    \Pi_{ab}=\frac{-\max({\mathcal F}, \bar {\mathcal B}_{ab})\alpha \bar c_{ab}\mu_{ab} + \beta \mu_{ab}^2}{\bar \rho_{ab}} \,, \ \text{for~~}{\bf v}_{ab}\cdot {\bf r}_{ab} < 0 \,,
    \label{eq:AV_turbulence}
\end{equation}
\noindent

\noindent with $0\le {\mathcal F} \le 1$ and $\bar {\mathcal B}_{ab}=\frac{1}{2}({\mathcal B}_{a}+{\mathcal B}_{b})$. The clamp reflects our goal of a general-purpose dissipation scheme: a floor value of $\mathcal F =1$ recovers the scheme used in previous tests. Allowing ${\mathcal F}\rightarrow 0$ yields excellent results in subsonic turbulence, but at the cost of effectively disabling AV in shear-dominated regions adjacent to shocks or in mixed regimes. We adopted the compromise of $\mathcal F =0.5$: AV is sufficiently attenuated for subsonic turbulence, yet does not vanish in, for instance, shock-dominated flows. According to Table~\ref{tab:table_cases_1}, this approach produces good results in all tests considered in this work.

Our metric to rank the turbulence models is to measure the location of the bottleneck peak ($k_{max}$) in the compensated velocity power spectrum. The works of \cite{val16} and \cite{cabezon2025} showed that SPH can reproduce both the inertial range and the bottleneck effect, provided that accurate gradients and sensitive AV control are used jointly. Therefore, given such a modern SPH implementation and a determined resolution, $k_{max}$ becomes a proxy for effective dissipation: without amplitude control of the AV, dissipation acts earlier (lower $k$) and the bottleneck intrudes sooner, effectively reducing the inertial range. Hence, a larger $k_{max}$ indicates a potential better model in this context.

Given the computational cost, we used the GPU-native SPH-EXA code to run at relatively high resolution, using $N=800^3$ particles. The domain is a box of a side $L=1$ with periodic boundary conditions and constant density $\rho=1$. Particles start from a relaxed glass-like distribution. The driving is done via an Ornstein-Uhlenbeck process \citep{ornstein_brownian_1930,federrath_statistics_2010}, following a parabolic amplitude field $\sim 1-(k-k_c)^2$, with $k=|\mathbf{k}|$, and $k_c=2$, which corresponds to a length of $L/2$, and vanishing for $k\leq 1$ and $k\geq 3$. This approach is implemented to reduce the impact of the stirring on the energy cascade, facilitating a natural transfer of energy from larger to smaller scales. Our forcing uses a natural mixing of $1/3$ compressive and $2/3$ solenoidal contributions. All simulations begin in a homogenous state and are stirred until they become statistically stationary at $t\simeq10$ with an RMS Mach number of $\mathcal{M}=0.3$. The times are normalized to the sound-crossing time, which in our setup is $t_{sc}=L/c=1$. We used a quasi-isothermal EOS with adiabatic index $\gamma=1.001$. This setup is provided out-of-the-box in SPH-EXA and can be reproduced by modifying the AV as desired and using the following command line:
\begin{center}
\resizebox{\columnwidth}{!}{\ttfamily
\$ ./sphexa-cuda --init turbulence --prop turbulence -n 800 -s 10.0 -w 1.0},
\end{center}

\noindent where \verb|-n| sets $N^{1D}$, \verb|-s| is the simulated physical time, and \verb|-w| is the output frequency.

\begin{figure}[h]
\centering
\includegraphics[width=0.50\textwidth]{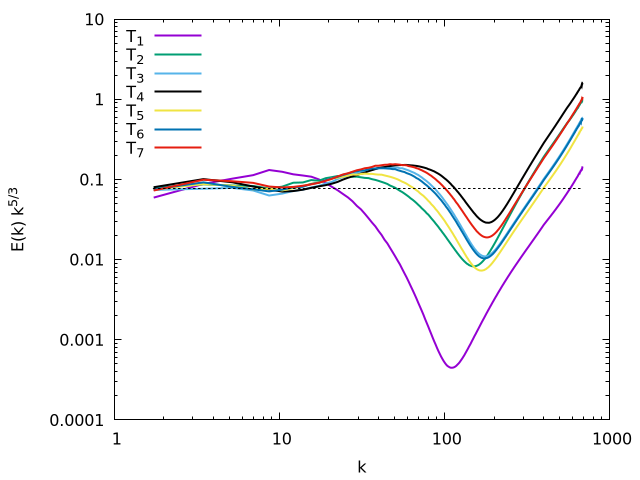}
\caption{Instantaneous compensated velocity power spectra at $t=10$.}
 \label{fig:turbulence}
\end{figure}

\begin{table}
        \centering
        \caption{Subsonic turbulence test: $k$ values for the maximum of the power spectra of the velocity field at the bottleneck.}
\begin{tabular}{cccccccc} 
    \hline
    & T$_1$ & T$_2$  & T$_3$  & T$_4$  & T$_5$ & T$_6$ & T$_7$\\ 
    \hline      
    $k_{max}$& 8.68 & 25.40 & 46.88 & 59.04 & 32.99 & 41.67 & 49.47\\
    \hline
\end{tabular}   
\label{tab:turbulence}
\end{table}

Figure~\ref{fig:turbulence} shows the resulting compensated power spectra at $t=10$, and Table~\ref{tab:turbulence} lists the corresponding $k_{max}$ values for models $T_k$ in Table~\ref{tab:table_cases}.
As expected, AV with constant coefficients (T$_1$) activates too early, pushing the bottleneck effect to low $k$, near the box scale, leaving no inertial plateau and quickly leading to viscous roll-off. The best model combines AVSW with SLR (T$_4$) yielding nearly a decade-wide plateau and bottleneck peaking at $k=59$. Pure SLR-based models, without AV amplitude modulation, achieve a very similar inertial range as (T$_4$), meaning that the large-to-intermediate dynamics are comparable. However, they show an earlier (and slightly stronger) encroachment of the bottleneck into the dissipation range. This is not surprising, since there is no mechanism to suppress AV in a smooth flow. Any residual compression or noise from particle disorder keeps a finite $\alpha$ active essentially all the time, leading to the earlier onset of dissipation. Model T$_7$ addresses this point, limiting the linear component of the AV, which delivers the second-best results.

\section{Searching for the optimal AV scheme}
\label{sec:optimalAV}
Our goal is this section is twofold: to classify the different AV schemes from our test results, and to discuss specific drawbacks that may affect methods that do not directly control the value of the parameter $\alpha$, such as SLR and Godunov-SPH.

\subsection{Ranking the AV schemes} 

\begin{table*}
\centering
\caption{Quantitative information used in Fig.~\ref{fig:best_to_worstmodels} to rank the first six tests cases with six AV schemes in Table~\ref{tab:table_cases}.}   
\begin{tabular}{c|c|c|c|c|c|c|c|c} 
\hline
\multirow{3}{*}{Test} & ST & \multicolumn{2}{c|}{S} & V & KHI & RTI & WC & T\\ 
& $\mathrm{L}_1 (v_x)$ & $\rho_{max}$ & $\mathrm {L}_1\ (v_{rad})$  & $\mathrm{L}_1 (v_{\phi})$& $\mathrm {A}(t/t_{KH}=1.5)$ & $\mathrm{KE_Y}$ & $\mathcal R_{1.8t_{KH}}$ & $k_{max}$\\
& $(\times 10^{-3})$ &  & $(\times 10^{-2})$ & $(\times 10^{-2})$ & $(\times 10^{-2})$ & $(\times 10^{-3})$ &  & \\
\hline
\hline
1-AV &2.97& 4.00  & 2.57 & 5.80 & 2.87 & 6.81& -0.105 & 8.68\\
2-AVSW  &5.90    & 4.02 &  2.69 & 2.54 & 8.31 & 8.49& -0.112 &  25.40  \\
3-AVSLR&3.16  &4.54 & 3.34 & 1.22 & 12.58 & 8.87& -0.139 & 46.88\\
4-AVSWSLR & 6.24   & 4.56 & 3.33 & 2.50 & 12.14& 8.39& -0.144 & 59.04\\
5-AVSLRB & 3.05  & 4.02 & 2.68 & 1.22 & 12.01& 8.62&-0.152 & 32.99\\
6-AVSLRB2 &3.04  & 4.03& 2.54 & 1.22& 12.52& 8.87&-0.162 & 41.67\\
\hline
7-AVSLRB2{$\mathcal F$}0.5 & 3.51  &4.07 &  2.78 & 1.39 & 12.29 & 8.50&-0.169 & 49.47  \\
\hline
\end{tabular}  
\tablefoot{The same information is also shown in the last row for a configuration giving good results in the turbulence study (see Sect.~\ref{sec:turbulence})}
\label{tab:table_cases_1}
\end{table*}
\begin{figure}[h]
\centering
\includegraphics[width=\columnwidth]{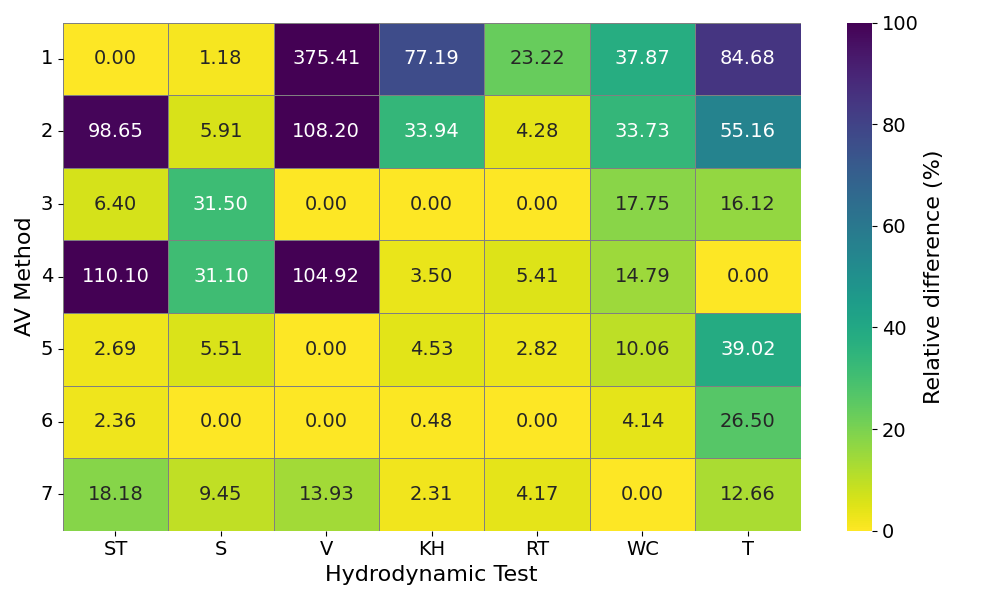}
\caption{Ranking heatmap of all AV methods for each test. Numbers show the relative difference to the best metric result overall for each test, using the values reported in Table~\ref{tab:table_cases_1}. Lower values indicate better results.}
 \label{fig:best_to_worstmodels}
\end{figure}
Given the spread of the results across the dissipation schemes in Table~\ref{tab:table_cases}, the best method is the one that is most balanced over the entire test suite. Generally speaking, for shocks, the best schemes maintain a high AV at the shock front (e.g., models S$_1$ and ST$_1$ in Table~\ref{tab:table_cases}). The opposite is true in regions subjected to strong shear, where AV should be kept low. On another note, the limitation in subsonic turbulence comes from the background random particle noise, which can spuriously trigger unnecessary AV. This needs to be controlled to allow for the emergence of the smaller-scale structure and the delay of the bottleneck and dissipation regimes.

Figure~\ref{fig:best_to_worstmodels} shows a ranking heatmap of AV variants across all tests, including turbulence. The heatmap is indicative and consistent with the case-by-case discussion above and summarized in Table \ref{tab:table_cases_1}. The x-axis lists the tests, while the y-axis lists the AV schemes, indexed as in Table~\ref{tab:table_cases_1}. Each cell is colored on a 0-100\% scale representing the relative distance to the best metric result on that test. Quantitative metrics are those reported in the previous sections: L$_1$ values (ST, S, V), the perturbation growth rate (KHI), the vertical kinetic energy (RTI), the cloud-destruction rate in a fixed time interval (WC), and the bottleneck peak wavenumber in the compensated velocity power spectra (T). In each column (test), the best results have a value of 0\%.

The horizontal accumulation of dark cells (many high percentages for a given method) in Fig.~\ref{fig:best_to_worstmodels} indicates a poor combination of methodologies, while the light yellow bands indicate a robust scheme. For example, the constant AV scheme (number 1) is good for handling shocks owing to its larger dissipation capability, but it is not suitable for handling shear flows and turbulence. Using switches, AVSW (number 2), improves over scheme 1, but remains weak on average compared to the other methods. The AVSLR scheme (number 3) is the opposite, excels in simulating shear flows, but is worse with shocks. This is in agreement with the results of \citet{sandnes2025}, but also underscores the need for extra care in shock capturing and treatment of turbulence. Mixing switches and SLR in the AVSWSLR scheme (number 4) yields the best result in turbulence  but degrades substantially in most other tests. The inclusion of the Balsara limiter to modulate the SLR, as suggested by \citet{sandnes2025}, AVSLRB (number 5), is able to cope with all fluid regimes successfully, except for turbulence. Our suggestion of using the square of the Balsara limiter in the SLR, AVSLRB2 (number 6) produces better results than method 5 and provides the best overall results in all tests, again with the exception of turbulence. Finally, adding the amplitude control of the AV, AVSLRB2$\mathcal{F}0.5$ provides a middle ground solution that behaves adequately overall, including subsonic turbulence.

\begin{table*}
        \centering
        \caption{Calculated models in the oscillating membrane test (Fig. \ref{fig:2D-wave}).}
\begin{tabular}{ccccccc} 
    \hline
    & AV & AVSW  & AVSLRB2  & AVSWSLR  & AVSLRB$^*2$ & AVSLRB$^*2\mathcal{F}05$\\ 
    \hline      
    $\alpha$ (min;max) & (0.2;0.2) & (0.05;1.0) & (1.0;1.0) & (0.05;1.0) & (1.0;1.0) & (0.5;1.0) \\
    modulation & - & - & $\mathcal{B}^2$& - & $\mathcal{B}{^*}^{2}$& $\mathcal{B}{^*}^{2}$\\
    \hline
\end{tabular}   
\tablefoot{Balsara numbers $\mathcal{B}$ and $\mathcal{B}^*$ are given by Eqs. \ref{eq:balsara_1} and \ref{eq:balsara_2}}
\label{tab:wavemembrane}
\end{table*}

\begin{figure}
\centering
\includegraphics[width=\columnwidth]{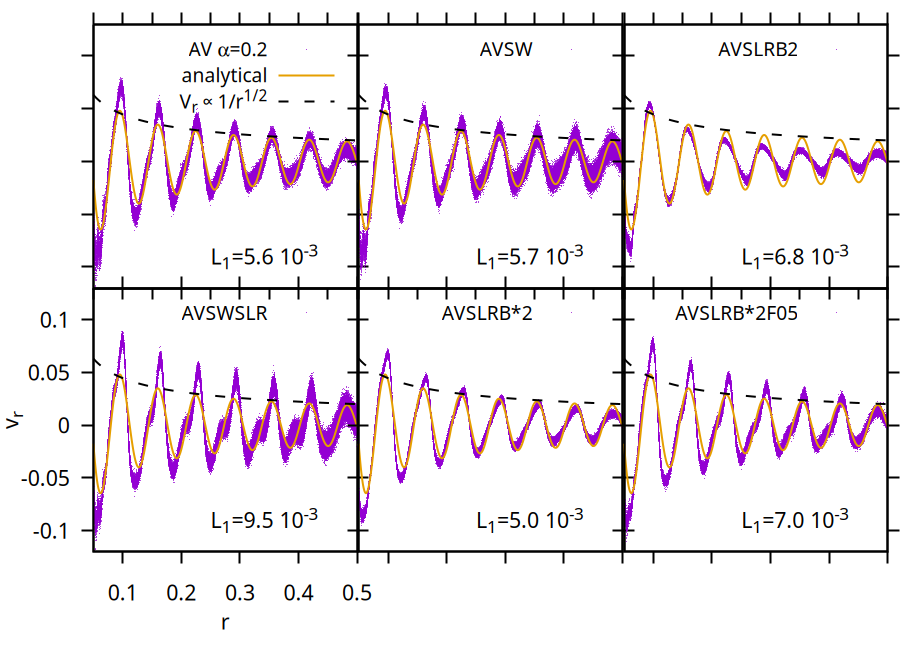}
\caption{Velocity profile of a surface wave at $t=0.35$ calculated with the AV  schemes in Table \ref{tab:wavemembrane}. The $L_1$ values obtained are also shown when compared to the analytical result. The $1/\sqrt{r}$ dependence of maximum amplitude with distance is also shown (dashed line).  }
    \label{fig:2D-wave}
\end{figure}
 
\begin{figure}
\centering
\includegraphics[width=0.49\textwidth]{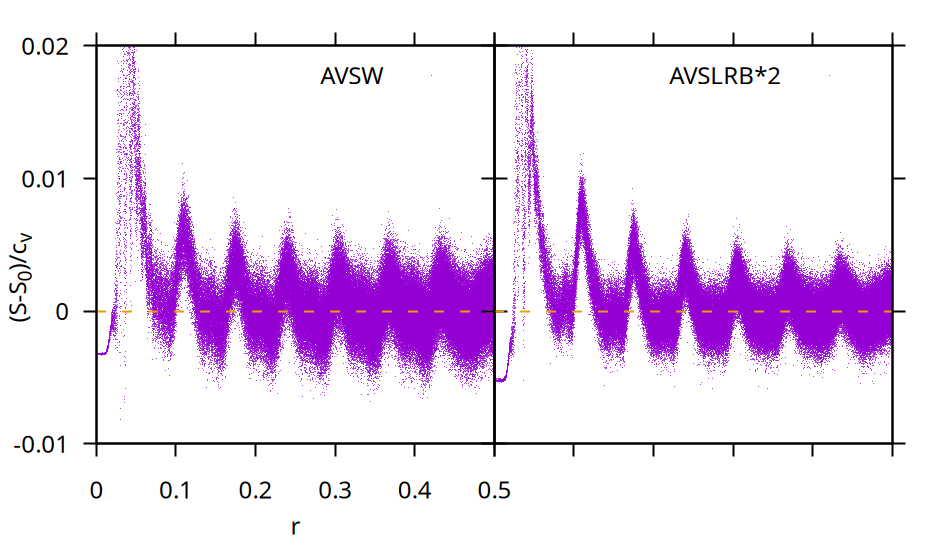}
\caption{Distribution of the entropy variation of each particle at $t=0.35$ for models AVSW and AVSLRB*2 in Table~\ref{tab:wavemembrane}}
    \label{fig:entropy_1}
\end{figure}

\begin{figure}
\centering
\includegraphics[width=0.46\textwidth]{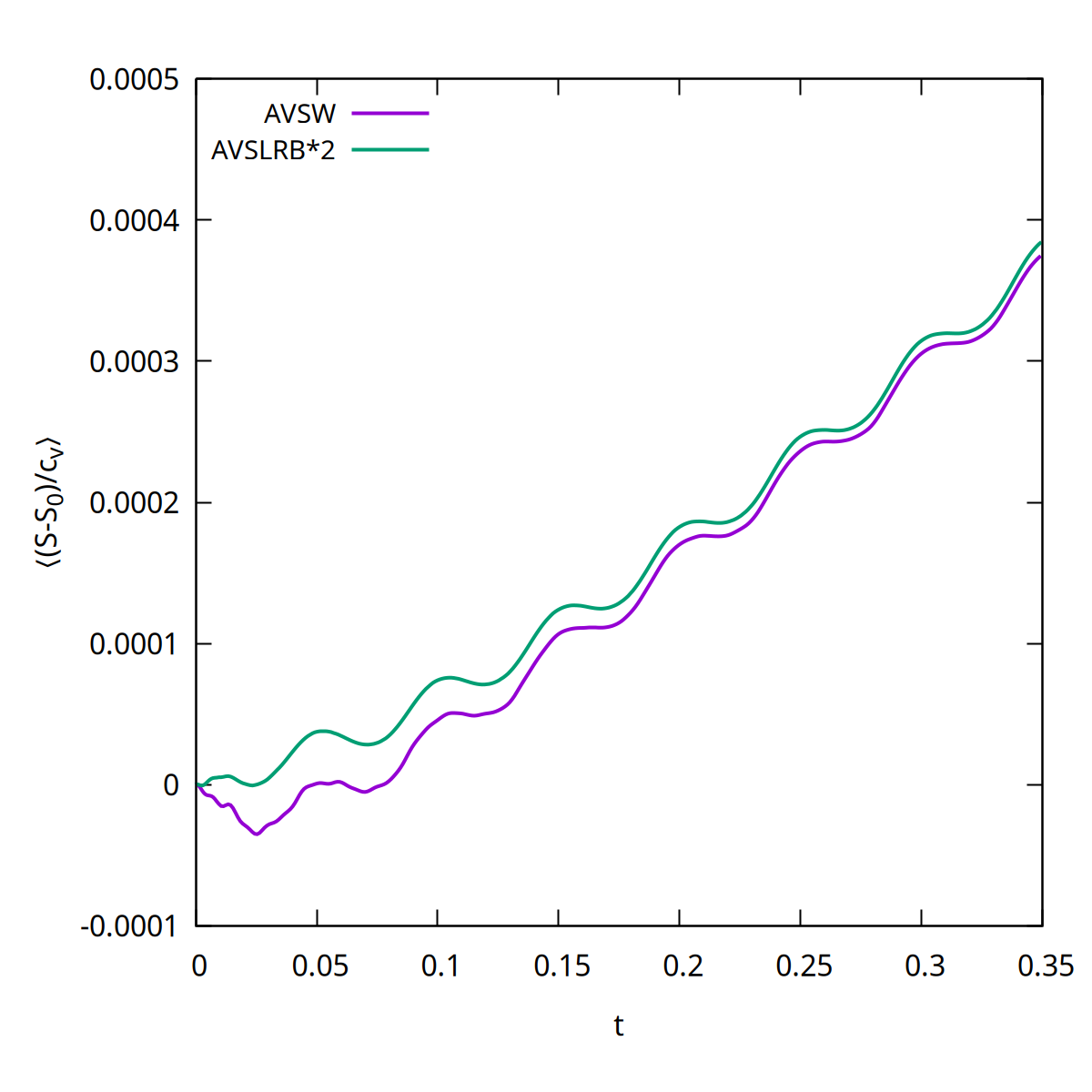}

    \caption{Evolution of averaged total entropy for models AVSW and AVSLRB*2 in Table \ref{tab:wavemembrane}}
    \label{fig:entropy_2}
\end{figure}

\subsection{Comments on possible disadvantages of SLR schemes}
\label{comments}

Despite the good performance of the AVSLR schemes shown above, two potential difficulties are worth noting. First, methods that do not explicitly control the AV coefficient $\alpha$ can struggle with linear-wave propagation. Second, in SLR-based formulations, the viscous form is not guaranteed to be strictly positive–definite pointwise, so care is needed to respect the second law of thermodynamics in discrete form.
 
Accurate propagation of nondissipative linear waves in SPH requires a negligible effective viscosity \citep{cul10}. Otherwise, the wave profile smears and its amplitude decays over time. \cite{puri2014a, puri2014b} analyzed alternatives to AV switches (e.g., Godunov–SPH) and found that such schemes can generally handle 1D linear waves less robustly than switch-based AV. An AV+SLR implementation could face similar issues: without an explicit, time-dependent control of $\alpha$, residual compression or noise might keep dissipation active. A straightforward remedy that involves reducing $\alpha$ aggressively will raise the noise level and might compromise other tests. A comprehensive study is beyond our scope, but this caveat deserves attention.

Astrophysical waves, however, are often surface or volume waves whose amplitude $A_m$ naturally decays with distance from the source as $A_m \propto r^{-1/2}$ and $A_m\propto r^{-1}$, respectively. This geometric decay can outweigh that due to excessive AV damping, making 2D/3D wavefields more forgiving than the 1D case. 

Here, we simulated low-amplitude waves generated by a harmonically oscillating cylindrical piston of a radius of $R_p=0.035$ and length  of $L=1$, with a period of $T=0.05$ and maximum radial velocity of $v^m_r = \pm 0.16$ \cite[see for details][]{garciasenz2012}. The box and particle number match the Sedov setup: initial $\rho=P=1$ give a sound speed of $c_s=1.29$ for $\gamma=5/3$. Although the amplitude of the wave is quite small, $\delta\rho/\rho\leq 6\%$, it is still considerably larger than that of typical sound waves and, therefore, a low amount of dissipation is expected. 

A complication is that the Balsara limiter, $\mathcal B$ (Eq.~\ref{eq:balsara_1}), used to modulate SLR, was not originally designed for wave propagation. In this problem, the oscillation of $\nabla\cdot {\bf{v}}$ with the wave period induces the oscillation of $\mathcal B$, with the result that $\mathcal{B}$ stays close to unity most of the time. Consequently, the SLRB and SLRB2 schemes behave similarly to the constant-$\alpha$ AV baseline (scheme 1 in Table~\ref{tab:av_variants}). Therefore, to make the limiter sensitive to oscillatory flows it is necessary to redefine the Balsara coefficient, adopting a frequency-aware variant. A simple way to achieve this is to consider 

\begin{equation}
{\mathcal B}^*_{a}= \frac{\vert\nabla\cdot \mathbf v\vert_a}{\vert\nabla\cdot \mathbf v\vert_a + \vert \nabla\times \mathbf v\vert_a+10^{-4}\frac{c_a}{h_a}+\frac{2\pi}{T_a}}\,, 
\label{eq:balsara_2}
\end{equation}

\noindent where $T$ is the known wave period (the piston period here). For high or low frequencies, we have $\mathcal{B}_a^*\rightarrow 0 / \mathcal{B}_a$, respectively.

Figure \ref{fig:2D-wave} compares the radial velocity $v_r(r)$ (with $r=\sqrt{x^2+y^2}$) for the AV variants listed in Table \ref{tab:wavemembrane} against the known analytic profile for waves propagating in a membrane \citep[e.g.,][]{elmore985}. We report the $L_1$ error in each panel. As expected, the lowest-viscosity cases (first, second, and fourth panels) show more noise but yield more symmetric waveforms. Switch-based AV exhibits larger scatter, though the mean follows the analytic profile well. The SLR model using the generalized limiter ${\mathcal B}^*$ (Eq.~\ref{eq:balsara_2}) shows a  comparable performance, with an even smaller $L_1$ error, although with a gradual loss of symmetry over time. This suggests that unlike strictly 1D linear waves (no geometric decay), diverging or converging waves in 2D or 3D can be represented reasonably well with AVSLRB*2 or its floored variant AVSLRB*2$\mathcal{F}05$, without any conflict with the rest of our test suite.

Finally, because SLR alters the reconstruction entering the dissipation term, we must ensure that the discrete viscous heating remains non-negative on average. As pointed out by \cite{price2020}, SLR schemes do not guarantee pointwise positive-definite AV by construction: the viscous work appears through products such as $({\bf v}_{a,i}-{\bf v}_{a,j})\cdot ({\bf v}'_{a,i}-{\bf v}'_{a,j}),$ where ${\bf v}'_a$ is the reconstructed velocity from Eq.~\ref{eq:filter_a}, and these products are not necessarily positive. This can lead to occasional local violations of the second law of thermodynamics; for example, as unphysical heating in rarefaction waves. In some Riemann solvers, entropy fixes can be introduced to mitigate such issues \citep{helluy2010}.  

In practice, the likehood of $({\bf v}_{a,i}-{\bf v}_{a,j})\cdot ({\bf v}'_{a,i}-{\bf v}'_{a,j})<0$ in shock-dominated systems is very small in our implementation. This is due to the Balsara modulation used in ${\bf v}'_a$ (Eq.~\ref{eq:filter_d}), which makes sign reversal unlikely for $\mathcal{B}\geq 0.5$. Furthermore, in the shock-tube test (Sect.~\ref{subsec:shock-tube}) we verified that no spurious entropy was produced in the rarefaction wave when using linear reconstruction. When $\mathcal {B}\lesssim 0.5$, the probability of sign reversals increases but still remains low. We monitored the fraction of negative pairwise products during the KHI run (model KH$_6$ in Table~\ref{tab:table_cases}), where we typically have $\mathcal{B}\lesssim 0.5$, and we found

\begin{equation}
    \frac {\mathrm {negative~ products}}{\mathrm {total~number~ of~ products}}\le 6\%\,,
\end{equation}

\noindent
with most inversions confined to small, noise-dominated velocities. 

In contrast, smooth subsonic flows with $\mathcal B\simeq 0$ are more prone to local, nonphysical entropy variations. We therefore tracked both the evolution of the entropy of each particle and the total entropy during the propagation of the waves produced by the oscillating piston described above. These waves are nearly linear, so we expect the total entropy to increase over time but by a small, almost marginal amount. The change in entropy of particle $a$ with respect to its initial value $S_{a,0}$ and normalized to the specific heat is 

\begin{equation}
    \left(\frac{S_a - S_{a,0}}{c_v}\right) = \ln\left(\frac{u_a}{u_{a,0}}\right)-(\gamma-1)\ln\left(\frac{\rho_a}{\rho_{a,0}}\right)
.\end{equation}

Figure \ref{fig:entropy_1} shows the distribution of $(S-S_0)/c_v$ at $t=0.35$ for models AVSW and AVSLRB*2. Far from the piston ($r > 0.1$), the entropy variation fluctuates almost symmetrically around zero in both cases, indicating a negligible change in entropy. This is consistent with the behavior of the total entropy shown in Fig.~\ref{fig:entropy_2} (excluding the piston region and normalized to the number of particles). We see that the evolution of $\left<\frac{S-S_0}{c_v}\right>$ is similar for both schemes and remains two orders of magnitude below the per-particle values shown in Fig.~\ref{fig:entropy_1}. Thus, summing these positive and negative local entropy variations yields a small, positive residual  (i.e. the expected slow increase of total entropy), with very similar trends in both calculations. We note that the AVSW curve shows a slight initial dip below zero, attributable to higher noise, before recovering.

\section{Conclusions}
\label{sec:conclusions}

 \cite{price2019} and \citet{sandnes2025} recently suggested that AV switches may be unnecessary if a slope-limited reconstruction (SLR) is used \citep[see also][]{fro17, rosswog2020, rosswog2020magma}. If confirmed, this assumption would generally represent a paradigm shift in the way shock-capturing and dissipation are  implemented in SPH codes. 

However, this does not come without a price. High, constant AV remains an excellent choice for shock capturing, although it does tend to overdamp turbulence, suppresses the emergence of hydrodynamic instabilities, and erase small-scale shear. Low, constant AV is great for turbulence, but produces strong post-shock oscillations. Switches may lag or trigger prematurely and leave residual noise in quiescent regions. Balsara limiters help in vortical flows, but they are unable to totally suppress unwanted dissipation in rotating shear flows. The bare SLR scheme reduces false compression triggers, but does not provide adequate dissipation in strong shocks and fails to describe them. In other words, there is always a trade-off to consider when choosing a specific approach. Moreover, a bare SLR is not enough to completely control spurious AV triggers by random particle movements in smooth flows and it does not completely ensure the correctness of the sign in the particles' entropy change. 

Even so, SLR is promising and, in our tests, it shows better results overall; in particular, for a Balsara-modulated SLR. In this work, we explore several SLR-based implementations across a broad test suite (shocks, shear flows, canonical instabilities, and subsonic turbulence) and suggest a balanced AV algorithm that performs robustly in all regimes. Our main findings are listed below. 

\begin{itemize}
     
    \item AVSLR schemes outperform our previous implementation of AVSW in SPHYNX across all tests, which is especially relevant for the study of subsonic turbulence. 

    \item In line with \citet{sandnes2025}, a good multi-regime performance requires modulating the SLR reconstruction with, for example, the Balsara limiter $\mathcal B$, as shown in Eqs.~(\ref{eq:filter_c}) and (\ref{eq:filter_d}). Although \citet{sandnes2025} used the exponent $p=1$, we find that $p=2$ (our AVSLRB2) consistently yields better results, particularly in turbulence, without sacrificing shock handling.

    \item Nonphysical entropy  behavior could appear if the bare SLR scheme is used (i.e., if SLR is not modulated). However, in regions subjected to  mild or strong dissipation, the $\mathcal {B}$-modulation of SLR works to inhibit any nonphysical sign reversal in the entropy.  Sign reversal can  still occur in noise-dominated regions where switches also fail (e.g., Fig.~\ref{fig:entropy_1}).

    \item Combining switches with SLR (AVSWSLR) tends to degrade shock and Gresho-Chan performance, but it works very well for subsonic turbulence, as shown in Fig.~\ref{fig:turbulence} and \cite{cabezon2025}. This improvement is due to the additional reduction of the AV parameter $\alpha$ introduced by the switches. 
    
    \item Motivated by the turbulence gains from reduced $\alpha$, we tested a simple amplitude control mechanism of AV in our AVSLRB2 scheme, considering $\alpha\rightarrow \alpha \times max[\mathcal F, \mathcal B]$, with $ 0\le {\mathcal F}\le 1$ in Eq.~\ref{eq:AV_turbulence}. Taking $\mathcal F=1$ recovers the AVSLRB2 scheme, while ${\mathcal F}=0.5$ improves the turbulence spectra, while producing only minor changes in the other tests (see the last two rows of Table~\ref{tab:table_cases_1} and Fig.~\ref{fig:best_to_worstmodels}). This offers a switch-free path to near-optimal turbulence behavior without compromising shocks or instabilities.

\end{itemize}

In short, AVSLRB2 (SLR with Balsara modulation at $p=2$) provides the best overall balance across shock-, shear-, and instability-dominated regimes. In addition, adding a modest AV amplitude floor (${\mathcal F}=0.5$) recovers much of the turbulence advantage seen with switches+SLR, without the broader downsides of switch-driven dissipation control. Therefore, we recommend using the AVSLRB2$\mathcal F$ scheme, setting the floor value to one in those cases where turbulence is not expected to be important, so that the scheme becomes AVSLRB2. In fluid regions where turbulence is relevant, it is more convenient to choose $\mathcal F =0.5$. 

An improvement of the technique would be to link the adopted value of $\mathcal F$ to the local Reynolds number of the flow. Another future line of development is to explore the impact of other limiters in the SLR, different from the one used in this work \citep{vanLeer1974}, since  \cite{rosswog2024} recently showed that changing the limiter can modify the level of dissipation. 

 Returning to the question posed in the title as to whether switches still necessary, our results suggest that for the regimes and resolutions explored here, time-dependent shock-trigger switches are likely not required to obtain robust dissipation control. In particular, SLR-based methods combined with continuous Balsara modulation (and, for subsonic turbulence, a modest amplitude floor) provide a single formulation that performs well from strong shocks to shear flows, instabilities, and low-Mach turbulence.

The results and conclusions presented here have been obtained using two state-of-the-art SPH codes, SPHYNX and SPH-EXA, together with a particular implementation of AV switches \citep{read2012}, which are also state-of-the-art schemes. Nonetheless, it would be valuable to repeat the same test suite with other modern SPH codes, such as SWIFT \citep{schaller2024} or PHANTOM \citep{price18}, which employ variants of the \citet{cul10} switch, to confirm (or further constrain) the advantages of SLR-based dissipation control over switch-based approaches.

\begin{acknowledgements}
We are very grateful to the referee, Daniel Price, both for constructive feedback and prompt reviewing of this work.
This work has been supported by the Swiss Platform for Advanced Scientific Computing (PASC) project "SPH-EXA: Optimizing Smoothed Particle Hydrodynamics for Exascale Computing". It has also been carried out as part of the SKACH consortium through funding from SERI. This work was also supported by a CHRONOS allocation from the Swiss National Supercomputing Centre (CSCS) under project ID ch18 on Alps. The authors acknowledge the support of the Center for Scientific Computing (sciCORE) at the University of Basel, where part of these calculations were performed.
\end{acknowledgements}


\begin{thebibliography}{54}
\expandafter\ifx\csname natexlab\endcsname\relax\def\natexlab#1{#1}\fi
\providecommand{\url}[1]{\texttt{#1}}
\providecommand{\path}[1]{#1}
\providecommand{\DOIprefix}{doi:}
\providecommand{\ArXivprefix}{arXiv:}
\providecommand{\URLprefix}{URL: }
\providecommand{\Pubmedprefix}{pmid:}
\providecommand{\doi}[1]{\href{http://dx.doi.org/#1}{\path{#1}}}
\providecommand{\Pubmed}[1]{\href{pmid:#1}{\path{#1}}}
\providecommand{\bibinfo}[2]{#2}
\ifx\xfnm\relax \def\xfnm[#1]{\unskip,\space#1}\fi

\bibitem[{{Agertz} {et~al.}(2007){Agertz}, {Moore}, {Stadel}, {Potter}, {Miniati}, {Read}, {Mayer}, {Gawryszczak}, {Kravtsov}, {Nordlund}, {Pearce}, {Quilis}, {Rudd}, {Springel}, {Stone}, {Tasker}, {Teyssier}, {Wadsley}, \& {Walder}}]{age07}
{Agertz}, O., {Moore}, B., {Stadel}, J., {et~al.} 2007, \mnras, 380, 963

\bibitem[{{Balsara}(1995)}]{bal95}
{Balsara}, D.~S. 1995, Journal of Computational Physics, 121, 357

\bibitem[{{Bauer} \& {Springel}(2012)}]{bauer2012}
{Bauer}, A. \& {Springel}, V. 2012, \mnras, 423, 2558

\bibitem[{{Cabez{\'o}n} \& {Garc{\'\i}a-Senz}(2024)}]{cabezon_mixedsincs_2024}
{Cabez{\'o}n}, R.~M. \& {Garc{\'\i}a-Senz}, D. 2024, Monthly Notices of the Royal Astronomical Society, 528, 3782

\bibitem[{{Cabez{\'o}n} {et~al.}(2012){Cabez{\'o}n}, {Garc{\'{\i}}a-Senz}, \& {Escart{\'{\i}}n}}]{cabezon2012}
{Cabez{\'o}n}, R.~M., {Garc{\'{\i}}a-Senz}, D., \& {Escart{\'{\i}}n}, J.~A. 2012, \aap, 545, A112

\bibitem[{{Cabez{\'o}n} {et~al.}(2017){Cabez{\'o}n}, {Garc{\'{\i}}a-Senz}, \& {Figueira}}]{cabezon2017}
{Cabez{\'o}n}, R.~M., {Garc{\'{\i}}a-Senz}, D., \& {Figueira}, J. 2017, \aap, 606, A78

\bibitem[{{Cabez{\'o}n} {et~al.}(2008){Cabez{\'o}n}, {Garc{\'{\i}}a-Senz}, \& {Rela{\~n}o}}]{cabezon2008}
{Cabez{\'o}n}, R.~M., {Garc{\'{\i}}a-Senz}, D., \& {Rela{\~n}o}, A. 2008, Journal of Computational Physics, 227, 8523

\bibitem[{{Cabez{\'o}n} {et~al.}(2025){Cabez{\'o}n}, {Garc{\'\i}a-Senz}, {Seckin Simsek}, {Keller}, {Sanz}, {Zhu}, {Mayer}, {Klessen}, \& {Ciorba}}]{cabezon2025}
{Cabez{\'o}n}, R.~M., {Garc{\'\i}a-Senz}, D., {Seckin Simsek}, O., {et~al.} 2025, arXiv e-prints, arXiv:2503.10273

\bibitem[{{Cavelan} {et~al.}(2020){Cavelan}, {Cabez{\'o}n}, {Grabarczyk}, \& {Ciorba}}]{cavelan2020}
{Cavelan}, A., {Cabez{\'o}n}, R.~M., {Grabarczyk}, M., \& {Ciorba}, F.~M. 2020, in PASC '20: Proceedings of the Platform for Advanced Scientific Computing ConferenceJune 2020, 11

\bibitem[{{Cha} {et~al.}(2010){Cha}, {Inutsuka}, \& {Nayakshin}}]{cha2010}
{Cha}, S.-H., {Inutsuka}, S.-I., \& {Nayakshin}, S. 2010, \mnras, 403, 1165

\bibitem[{{Cha} \& {Whitworth}(2003)}]{cha03}
{Cha}, S.-H. \& {Whitworth}, A.~P. 2003, \mnras, 340, 73

\bibitem[{{Cullen} \& {Dehnen}(2010)}]{cul10}
{Cullen}, L. \& {Dehnen}, W. 2010, \mnras, 408, 669

\bibitem[{{Elmore} \& {Heald}(1985)}]{elmore985}
{Elmore}, W.~C. \& {Heald}, M.~A. 1985, {Physics of Waves} (Dover Publications, Inc., New York)

\bibitem[{{Federrath} {et~al.}(2010){Federrath}, {Roman-Duval}, {Klessen}, {Schmidt}, \& {Mac Low}}]{federrath_statistics_2010}
{Federrath}, C., {Roman-Duval}, J., {Klessen}, R.~S., {Schmidt}, W., \& {Mac Low}, M.~M. 2010, Astronomy \& Astrophysics, 512, A81

\bibitem[{{Frontiere} {et~al.}(2017){Frontiere}, {Raskin}, \& {Owen}}]{fro17}
{Frontiere}, N., {Raskin}, C.~D., \& {Owen}, J.~M. 2017, Journal of Computational Physics, 332, 160

\bibitem[{{Garc{\'{\i}}a-Senz} {et~al.}(2012){Garc{\'{\i}}a-Senz}, {Cabez{\'o}n}, \& {Escart{\'{\i}}n}}]{garciasenz2012}
{Garc{\'{\i}}a-Senz}, D., {Cabez{\'o}n}, R.~M., \& {Escart{\'{\i}}n}, J.~A. 2012, \aap, 538, A9

\bibitem[{{Garc{\'\i}a-Senz} {et~al.}(2022){Garc{\'\i}a-Senz}, {Cabez{\'o}n}, \& {Escart{\'\i}n}}]{garciasenz2022}
{Garc{\'\i}a-Senz}, D., {Cabez{\'o}n}, R.~M., \& {Escart{\'\i}n}, J.~A. 2022, \aap, 659, A175

\bibitem[{{Gingold} \& {Monaghan}(1977)}]{gingold77}
{Gingold}, R.~A. \& {Monaghan}, J.~J. 1977, \mnras, 181, 375

\bibitem[{Godunov(1959)}]{godunov59}
Godunov, S.-K. 1959, Sb. Math., 47

\bibitem[{{Gresho} \& {Chan}(1990)}]{gresho1990}
{Gresho}, P.~M. \& {Chan}, S.~T. 1990, International Journal for Numerical Methods in Fluids, 11, 621

\bibitem[{{Harten} {et~al.}(1976){Harten}, {Hyman}, \& {Lax}}]{harten1976}
{Harten}, A., {Hyman}, J.~M., \& {Lax}, P.~D. 1976, Communications in Pure Applied Mathematics, 29, 297

\bibitem[{{Helluy} {et~al.}(2010){Helluy}, {H{\'e}rard}, {Mathis}, \& {M{\"u}ller}}]{helluy2010}
{Helluy}, P., {H{\'e}rard}, J.-M., {Mathis}, H., \& {M{\"u}ller}, S. 2010, Comptes Rendus Mecanique, 338, 493

\bibitem[{{Inutsuka}(2002)}]{inu02}
{Inutsuka}, S.-I. 2002, Journal of Computational Physics, 179, 238

\bibitem[{Keller {et~al.}(2023)Keller, Cavelan, Cabezon, Mayer, \& Ciorba}]{keller2023}
Keller, S., Cavelan, A., Cabezon, R., Mayer, L., \& Ciorba, F. 2023, in Proceedings of the Platform for Advanced Scientific Computing Conference, PASC '23 (New York, NY, USA: Association for Computing Machinery)

\bibitem[{{Lucy}(1977)}]{lucy77}
{Lucy}, L.~B. 1977, \aj, 82, 1013

\bibitem[{{McNally} {et~al.}(2012){McNally}, {Lyra}, \& {Passy}}]{mcn12}
{McNally}, C.~P., {Lyra}, W., \& {Passy}, J.-C. 2012, \apjs, 201, 18

\bibitem[{{Monaghan}(1992)}]{mon1992}
{Monaghan}, J.~J. 1992, \araa, 30, 543

\bibitem[{{Monaghan}(1997)}]{mon1997}
{Monaghan}, J.~J. 1997, Journal of Computational Physics, 136, 298

\bibitem[{{Monaghan}(2005)}]{mon2005}
{Monaghan}, J.~J. 2005, Reports on Progress in Physics, 68, 1703

\bibitem[{{Monaghan} \& {Gingold}(1983)}]{mon1983}
{Monaghan}, J.~J. \& {Gingold}, R.~A. 1983, Journal of Computational Physics, 52, 374

\bibitem[{{Morris} \& {Monaghan}(1997)}]{morris1997}
{Morris}, J.~P. \& {Monaghan}, J.~J. 1997, Journal of Computational Physics, 136, 41

\bibitem[{{Osher} \& {Chakravarthy}(1984)}]{osher1984}
{Osher}, S. \& {Chakravarthy}, S. 1984, SIAM Journal on Numerical Analysis, 21, 955

\bibitem[{{Price}(2008)}]{price08}
{Price}, D.~J. 2008, Journal of Computational Physics, 227, 10040

\bibitem[{Price(2012)}]{price12}
Price, D.~J. 2012, \mnras, 420, L33

\bibitem[{{Price}(2024)}]{price2019}
{Price}, D.~J. 2024, arXiv e-prints, arXiv:2407.10176

\bibitem[{{Price} \& {Laibe}(2020)}]{price2020}
{Price}, D.~J. \& {Laibe}, G. 2020, \mnras, 495, 3929

\bibitem[{{Price} {et~al.}(2018){Price}, {Wurster}, {Tricco}, {Nixon}, {Toupin}, {Pettitt}, {Chan}, {Mentiplay}, {Laibe}, {Glover}, {Dobbs}, {Nealon}, {Liptai}, {Worpel}, {Bonnerot}, {Dipierro}, {Ballabio}, {Ragusa}, {Federrath}, {Iaconi}, {Reichardt}, {Forgan}, {Hutchison}, {Constantino}, {Ayliffe}, {Hirsh}, \& {Lodato}}]{price18}
{Price}, D.~J., {Wurster}, J., {Tricco}, T.~S., {et~al.} 2018, \pasa, 35, e031

\bibitem[{{Puri} \& {Ramachandran}(2014{\natexlab{a}})}]{puri2014a}
{Puri}, K. \& {Ramachandran}, P. 2014{\natexlab{a}}, Journal of Computational Physics, 256, 308

\bibitem[{{Puri} \& {Ramachandran}(2014{\natexlab{b}})}]{puri2014b}
{Puri}, K. \& {Ramachandran}, P. 2014{\natexlab{b}}, Journal of Computational Physics, 270, 432

\bibitem[{{Read} \& {Hayfield}(2012)}]{read2012}
{Read}, J.~I. \& {Hayfield}, T. 2012, \mnras, 422, 3037

\bibitem[{{Rosswog}(2015)}]{ros15}
{Rosswog}, S. 2015, \mnras, 448, 3628

\bibitem[{{Rosswog}(2020{\natexlab{a}})}]{rosswog2020}
{Rosswog}, S. 2020{\natexlab{a}}, \apj, 898, 60

\bibitem[{{Rosswog}(2020{\natexlab{b}})}]{rosswog2020magma}
{Rosswog}, S. 2020{\natexlab{b}}, \mnras, 498, 4230

\bibitem[{{Rosswog}(2024)}]{rosswog2024}
{Rosswog}, S. 2024, arXiv e-prints, arXiv:2411.19228; accepted for publication in Computer modeling in Engineering \& Sciences 2025 143(2) 1713.

\bibitem[{{Rosswog} \& {Diener}(2025)}]{rosswog_diener2025}
{Rosswog}, S. \& {Diener}, P. 2025, in New Frontiers in GRMHD Simulations, ed. C.~{Bambi}, Y.~{Mizuno}, S.~{Shashank}, \& F.~{Yuan}, 235--273

\bibitem[{{Sandnes} {et~al.}(2025){Sandnes}, {Eke}, {Kegerreis}, {Massey}, {Ruiz-Bonilla}, {Schaller}, \& {Teodoro}}]{sandnes2025}
{Sandnes}, T.~D., {Eke}, V.~R., {Kegerreis}, J.~A., {et~al.} 2025, Journal of Computational Physics, 532, 113907

\bibitem[{{Schaller} {et~al.}(2024){Schaller}, {Borrow}, {Draper}, {Ivkovic}, {McAlpine}, {Vandenbroucke}, {Bah{\'e}}, {Chaikin}, {Chalk}, {Chan}, {Correa}, {van Daalen}, {Elbers}, {Gonnet}, {Hausammann}, {Helly}, {Hu{\v{s}}ko}, {Kegerreis}, {Nobels}, {Ploeckinger}, {Revaz}, {Roper}, {Ruiz-Bonilla}, {Sandnes}, {Uyttenhove}, {Willis}, \& {Xiang}}]{schaller2024}
{Schaller}, M., {Borrow}, J., {Draper}, P.~W., {et~al.} 2024, \mnras, 530, 2378

\bibitem[{{Sedov}(1959)}]{sedov1959}
{Sedov}, L.~I. 1959, {Similarity and Dimensional Methods in Mechanics}

\bibitem[{{Sod}(1978)}]{sod1978}
{Sod}, G.~A. 1978, Journal of Computational Physics, 27, 1

\bibitem[{Uhlenbeck \& Ornstein(1930)}]{ornstein_brownian_1930}
Uhlenbeck, G.~E. \& Ornstein, L.~S. 1930, Phys. Rev., 36, 823

\bibitem[{{Valdarnini}(2016)}]{val16}
{Valdarnini}, R. 2016, \apj, 831, 103

\bibitem[{{van Leer}(1974)}]{vanLeer1974}
{van Leer}, B. 1974, Journal of Computational Physics, 14, 361

\bibitem[{{Wadsley} {et~al.}(2017){Wadsley}, {Keller}, \& {Quinn}}]{Wadsley17}
{Wadsley}, J.~W., {Keller}, B.~W., \& {Quinn}, T.~R. 2017, \mnras, 471, 2357

\bibitem[{{Yee} {et~al.}(1983){Yee}, {Warming}, \& {Harten}}]{yee1983}
{Yee}, H.~C., {Warming}, R.~F., \& {Harten}, A. 1983, {Implicit Total Variation Diminishing (TVD) schemes for steady-state calculations}, NASA Technical Memorandum, NASA/STI Accession number: 19830014814

\end{thebibliography}

\end{document}